\documentclass[12pt]{article}
\usepackage{fullpage}
\usepackage{graphicx}
\usepackage[nil]{babel}
\usepackage{amsmath}
\usepackage{amsfonts}
\usepackage{natbib}
\usepackage{url}
\usepackage{enumitem}
\usepackage[doublespacing]{setspace}
\usepackage{booktabs}
\usepackage{tabularx}
\usepackage{multirow}
\usepackage{arydshln}
\usepackage{color}
\usepackage{textcomp}
\usepackage{mathtools}
\usepackage{titling}
\usepackage{rotating}

\graphicspath{{./paper/}{./paper_new/}}
\title{\bf Zero-inflated Poisson Factor Model with Application to Microbiome Absolute Abundance Data}

\author{Tianchen Xu$^{*}$\\
tx2155@columbia.edu\\
Department of Biostatistics, Columbia University, NY 10032, USA\\[10pt]
Ryan T. Demmer\\
demm0009@umn.edu\\
Division of Epidemiology, University of Minnesota, MN 55454, USA\\[10pt]
Gen Li\\
gl2521@cumc.columbia.edu\\
Department of Biostatistics, Columbia University, NY 10032, USA}

\begin{document}
\sloppy

\newcommand{\e}{\operatorname{E}}
\newcommand{\prodn}{\prod_{i=1}^n}
\newcommand{\sumn}{\sum_{i=1}^n}
\newcommand{\dott}{$\,\boldsymbol{^\cdot}$}
\newcommand{\dottt}{$\,^\star$}
\newcommand{\dotttt}{$\,^{\star\star}$}
\newcommand{\dottttt}{$\,^{\star\star\star}$}
\maketitle

\newpage
\begin{abstract}
Dimension reduction of high-dimensional microbiome data facilitates subsequent analysis such as regression and clustering. Most existing reduction methods cannot fully accommodate the special features of the data such as count-valued and excessive zero reads.  We propose a zero-inflated Poisson factor analysis (ZIPFA) model in this article. The model assumes that microbiome absolute abundance data follow zero-inflated Poisson distributions with library size as offset and Poisson rates negatively related to the inflated zero occurrences. The latent parameters of the model form a low-rank matrix consisting of interpretable loadings and low-dimensional scores which can be used for further analyses. We develop an efficient and robust expectation-maximization (EM)  algorithm for parameter estimation. We demonstrate the efficacy of the proposed method using comprehensive simulation studies. The application to the Oral Infections, Glucose Intolerance and Insulin Resistance Study (ORIGINS) provides valuable insights into the relation between subgingival microbiome and periodontal disease.
\end{abstract}
keywords: 16S sequencing; Factor analysis; Low rank; Microbiome data; Zero inflation.

\newpage

\section{Introduction}
The development of next-generation sequencing (NGS) technologies enables the quantification of microbes living in and on the human body
\citep{hamady2009microbial}.
Many recent studies have identified that microbial dysbiosis in specific anatomical sites is associated with complex diseases such as type 2 diabetes, prediabetes, insulin resistance and cardiovascular disease
\citep{demmer2015periodontal, demmer2017subgingival, dewhirst2010human}.
However, detecting microbial association remains a formidable problem due to the complexity of microbiome data and a lack of appropriate  statistical methods
\citep{li2015microbiome}.

In the motivating Oral Infections, Glucose Intolerance and Insulin Resistance Study (ORIGINS), one goal is to investigate the relationship between subgingival microbial communities and both periodontal disease and biomarkers of diabetes risk. Previous analysis of NGS data in ORIGINS shows expected associations between oral bacterial phyla and markers of inflammation and impaired glucose regulation. However, at the taxa level, few individual bacterial taxa are identified with statistical significance
\citep{demmer2017subgingival}
limiting the ability to understand which taxa drive phylum level findings.
One of the challenges arises from the high dimensionality of data, requiring multiple testing corrections that results in reduced statistical power. Thus, dimension reduction is often desired to reduce the number of variables subject to hypothesis testing. There are several outstanding challenges for dimension reduction of microbiome NGS data:
1) Library sizes are heterogeneous across samples. 2) Typical data in a microbiome study consist of highly skewed non-negative sequence counts
\citep{hamady2009microbial},
which cannot be directly modeled with Gaussian distributions
\citep{srivastava2010two}.
3) The data contain excessive zeros. There are two types of zero counts in microbiome data: one is ``true zeros'' (i.e., absence of taxa in samples) and the other is ``pseudo zeros'' (i.e., the presence is below detection limit). Either true absence or undetected presence of a taxon will lead to excessive zeros in microbiome data.




%

Factor analysis has been widely used to identify low-dimensional features in high dimensional data.
Typically, original read counts are first converted to compositions (or rarefied) and then transformed. Standard factor analysis is applied to the transformed data to achieve dimension reduction
\citep{mcmurdie2014waste}.
However, there is significant information loss  during the preprocessing step, and the compositional data are still difficult to handle statistically because of the extra constraint on the sum and excessive zeros.
\citet{mcmurdie2014waste}
pointed out that both preprocessing approaches are inappropriate for detection of differentially abundant species. Therefore, standard factor analysis methods are not adequate for analyzing microbiome absolute abundance (i.e., sequence read counts) and thus it is desired to bypass the preprocessing procedure and model absolute abundance data directly.

There are some recent developments on modeling sequence read count data.
\citet{lee2013poisson}
developed a Poisson factor model with offset. This method can effectively model count-valued data with heterogeneous library sizes, but it fails to take the excessive zeros into consideration.
\citet{cao2017microbial}
developed a Poisson-Multinomial model to model the high variation arising from excessive zeros. However, the method is not adequate to address the over-dispersion from excessive zeros in data
\citep{li2015microbiome}. In order to account for this extra biological variability, established statistical theory shows using a mixture model is necessary
\citep{mcmurdie2014waste,lu2005identifying}.
Very recently,
\citet{b2018glm}
built a zero-inflated quasi-Poisson factor model to conquer the inflated zero challenge, but this model relies on a somewhat unrealistic assumption that each taxon has a fixed zero probability despite the heterogeneity in samples. It does not establish any link between the probability of true zeros and the Poisson  rate underlying the microbiome absolute abundances, which might lead to inferior results as we will show later.

\begin{figure}[!t]
	\centering\includegraphics[width=0.75\textwidth]{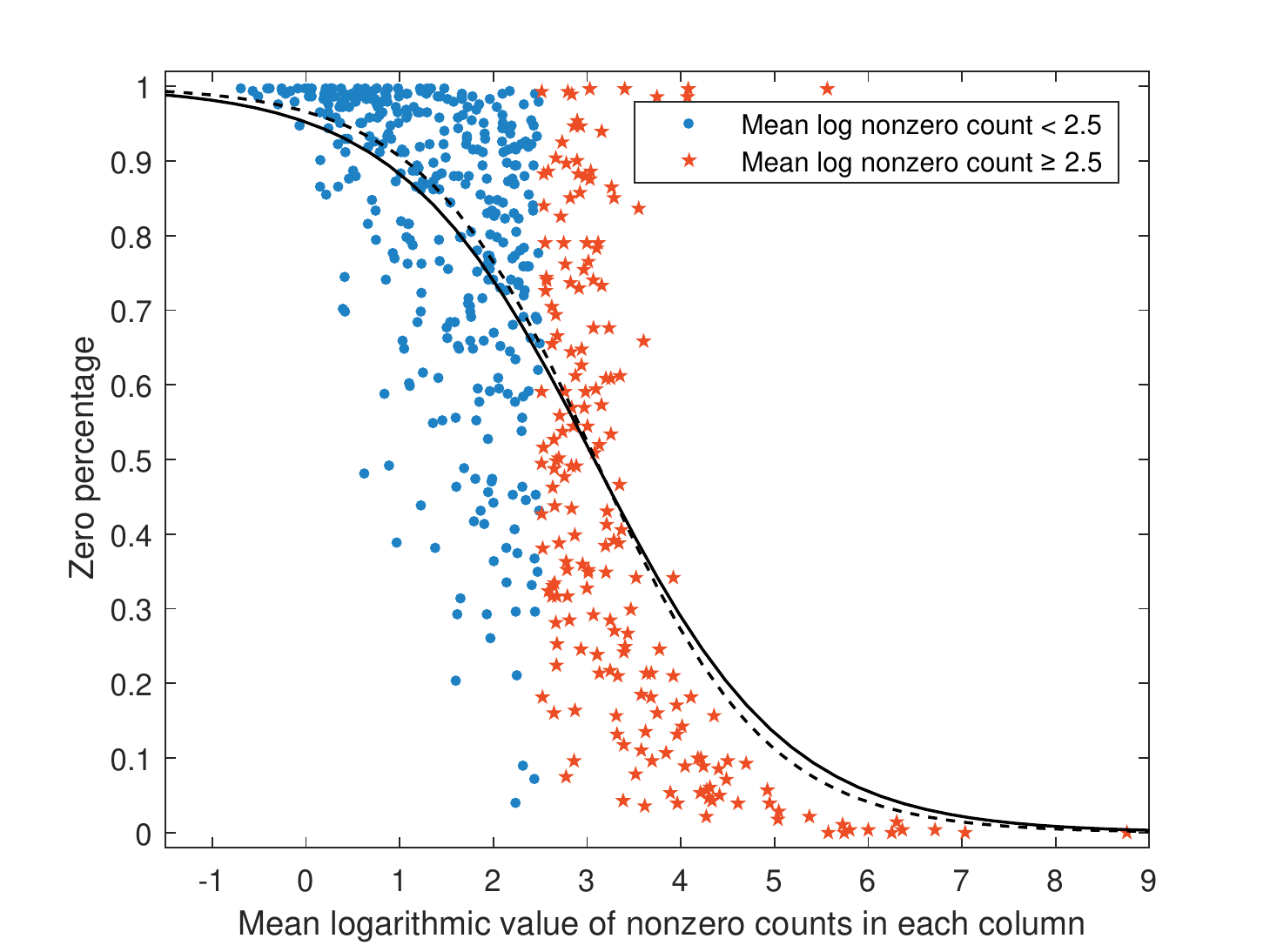}
		\caption{Relationship between zero probability and counts value. Zero percentage decreases for taxa that have higher mean logarithmic nonzero counts. The red stars are the columns that have mean log nonzero counts large than $2.5$ and the blue dots are other columns. The dashed curve is the fitted logistic function for red stars and the solid curve is the fitted logistic function for all points. }
		\label{p_proportion}
\end{figure}

Intuitively, the probability for a count value being true zero should be lower if the underlying Poisson rate is relatively high.
\citet{cao2017microbial}
also presented similar finding when analyzing gut microbiome data. In order to further validate this conjecture, in our motivating ORIGINS data, we plot the mean logarithmic nonzero count of each taxon against zero percentage of the same taxon in Figure~\ref{p_proportion}.
Taxa with large mean log values (i.e., greater than $2.5$) are marked as red stars, and those with low mean log values are marked as blue dots. The observed zero percentage of red stars should approximate the true zero probability, as it is unlikely to have zeros from a Poisson distribution with large rate ($\Pr\{X=0\}=5.1\times 10^{-6}$ for $X\sim Poisson(\exp(2.5))$). We observe that, for those red stars,  the zero percentage is lower in taxa with larger mean values of log counts, reflecting  the negative relationship between the probability of ``true" zeros and the underlying Poisson rate.
We further fit logistic curves to the red stars (dashed curve) and all data points (solid curve), respectively. The two curves are very close to each other, indicating the whole data set follows a zero generation mechanism similar to that for the red stars. To further characterize the relationship, we generate a simulated data set with the same library size, total zero proportion and mean nonzero count for each taxon as in the real data, but set the probability of true zeros to be independent of the underlying Poisson rate. In Web Figure~1, the observed zero percentage decreases first and then levels off at a fixed value, which is very different from the real data pattern in Figure~\ref{p_proportion}. In addition, when we partition the taxa based on the mean log values as before and fit two logistic curves, one for the red stars and the other for all data points, the fit is very poor and the two curves are very different. This result correctly indicates there is no relationship between the probability of ``true" zeros and the underlying Poisson rates in this simulation example. Such findings support the conjecture about the negative relation between the true zero probability and Poisson rate, and motivates us to adopt a parametric link between the two in Section~\ref{method}.

In summary, we develop a new zero-inflated Poisson factor analysis (ZIPFA) model for reducing the dimension of microbiome data. The model has unique contributions to microbiome data analysis:
\begin{itemize}[parsep=0pt,itemsep=0pt,topsep=0pt,labelindent=\parindent,leftmargin=*]
	\item It properly models the original absolute abundance count data;
	\item It specifically accommodates excessive zero counts in data;
	\item It incorporates a realistic link between the true zero probability and Poisson rate.
\end{itemize}

The rest of the article is organized as follow. In Section~\ref{method}, we introduce the proposed ZIPFA model first, and then discuss the  fitting algorithm and rank selection in the last part of the section. In Section~\ref{sec:simulation}, we present simulation studies to compare different methods. The analysis of ORIGINS study and results are in Section~\ref{sec:application}. Finally, We conclude the article with discussions in Section~\ref{sec:dicussion}.

\section{Method}\label{method}
\subsection{Model Setup}\label{sec:model_setup}

In microbiome studies, the absolute sequencing read counts are summarized in a matrix $A \in \mathbb{N}_0^{n\times m}$, where $n$ is the sample size and $m$ is the number of taxa. Let $A_{ij}$ represents the read count of taxon $j$ of individual $i$ $(i=1,\cdots,n;\, j=1,\cdots,m)$. Let $N = (N_1,N_2,\cdots,N_n)^\top$ a be vector of the relative library sizes where:
\begin{equation*}
	N_i =\sum_{j=1}^m A_{ij}\bigg/\operatorname{median}\left( \sum_{j=1}^m A_{1j},\; \sum_{j=1}^m A_{2j},\; \cdots,\; \sum_{j=1}^m A_{nj} \right).
\end{equation*}
Unlike the absolute library size which depends on the number of measured taxa and sequence depth, the relative library size is scale invariant and performs favorably in numerical studies.

Since excessive zeros may come from true absence or undetected presence of taxa, a mixed distribution is proper to describe $A_{ij}$
\citep{b2018glm}.
Previous research points out that it is reasonable to assume each read count $A_{ij}$ follows a zero-inflated Poisson (ZIP) distribution
\citep{xu2015assessment}:
\begin{equation*}
	A_{ij}\sim
	\begin{dcases*}
	0, & with prob $= p_{ij}$\\
	Poisson(N_i\lambda_{ij}), & with prob $= 1-p_{ij}$
	\end{dcases*}
\end{equation*}
where $p_{ij}\; (0\le p_{ij} \le 1)$ is the unknown parameter of the Bernoulli distribution that describes the occurrence of true zeros; $\lambda_{ij}\; (\lambda>0)$ is the unknown parameter of the normalized Poisson part, and $N_i \lambda_{ij}$ is the Poisson rate adjusted by the subject-specific relative library size $N_{i}$. Then let $P = \operatorname{logit}(p_{ij}) \in \mathbb{R}^{n\times m}$ and  $\Lambda = \ln( \lambda_{ij})\in \mathbb{R}^{n\times m}$ be the corresponding natural parameter matrices to map parameters $p_{ij}$, $\lambda_{ij}$ to the real line.

To link the negative relationship between true zero probability $p_{ij}$ and Poisson rate $\lambda_{ij}$ in Figure~\ref{p_proportion}, we propose to use a positive shape parameter $\tau$ to build the logistic link by modeling $P=-\tau \Lambda$ (i.e.,  $\operatorname{logit}(p_{ij})=-\tau \ln(\lambda_{ij})$). In the setting, $\ln(\lambda_{ij}) $ and $\operatorname{logit}(p_{ij})$ are the natural links that linearize the normalized Poisson mean and the Bernoulli probability of true zeros. The probability of $A_{ij}$ being true zero decreases when $\lambda_{ij}$ increases.

To encourage dimension reduction, we assume matrix $\Lambda \in \mathbb{R}^{n\times m}$ has a low rank structure $\Lambda = UV^\top$ with rank $K<\min(m, n)$, where $U\in \mathbb{R}^{n\times K}$ is the score matrix; $V \in \mathbb{R}^{m\times K}$ is the loading matrix. Then the proposed ZIPFA model with rank $K$ is given by:
\begin{equation*}
		\begin{dcases}
A_{ij}\sim \text{ZIP distribution}\\
\operatorname{logit}(p_{ij})=-\tau \ln(\lambda_{ij}) \\
\ln (\lambda_{ij})=u_{i1}v_{j1}+u_{i2}v_{j2}+\cdots+u_{iK}v_{jK}
\end{dcases}
\end{equation*}
where $u_{ij}$, $v_{ij}$ are elements of $U$, $V$. Here, $u_{ij}$ represents the j\textsuperscript{th} factor score for the i\textsuperscript{th} individual, and
$v_{ij}$ is the i\textsuperscript{th} taxon loading on j\textsuperscript{th} factor.

Once the factor number $K$ is determined and  $U$, $V$ matrix are estimated, we  reduce the dataset dimension from $m$ to $K$. Score matrix $U$ contains the same sample size as the original dataset $A$ but only $K$ variables. It is much easier to associate the clinical outcomes with the low dimensional score matrix $U$ through  regression analysis. The loading matrix $V$ reflects  the composition of the factors in $U$. Each column in $V$ corresponds to a factor in $U$ and their values show the importance of original taxa in the corresponding factors.\\

\begin{minipage}{0.98\textwidth}
	\noindent
	\begin{tabularx}{\textwidth}{X}
		\toprule
		\multicolumn{1}{c}{Algorithm 1: The ZIPFA algorithm}                    \\
		\hline
		Matrix $A \in \mathbb{R}^{n\times m}$ is to be decomposed to $K$ factors. \\
		\textbf{Initialize:}
		\begin{enumerate}[labelindent=\parindent,leftmargin=*]
			\item  Let $\tilde{A}$ be the same matrix as $A$ but all $0$'s are replaced by the column mean;
			\item  Apply SVD to $\ln(\tilde{A})$ to obtain the components $U^{old}$ and $V^{old}$.
		\end{enumerate}
		\textbf{Update:}
		\begin{enumerate}[labelindent=\parindent,leftmargin=*]
			\setcounter{enumi}{2}
			\item Fit zero-inflated Poisson regression with $A^{(u)}$ as the response, $U^{\star\,old}$ as the covariates and $N^{(u)}$ as the scaling vector to obtain the estimated $V^{new}$;\label{itm:solve_u}
			\item Fit zero-inflated Poisson regression with $A^{(v)}$ as the response, $V^{\star\,new}$ as the covariates and $N^{(v)}$ as the scaling vector to obtain the estimated $U^{new}$ ;\label{itm:solve_v}
			\item Apply SVD to $U^{new}V^{new\,T}$ and obtain $U^{old}$ and $V^{old}$;\label{itm:svd}
			\item Repeat from step 3 until convergence.
		\end{enumerate}\\
		\bottomrule
	\end{tabularx}
\end{minipage}\\

\subsection{Maximum-likelihood Estimation}\label{sec:max_like}
To begin our discussion, some notation needs to be introduced. Let row vectors $a_{(i,)}$, $u_{(i,)}$, $v_{(i,)}$ denote the i\textsuperscript{th} row of $A$, $U$, $V$. Let column vectors $a_{(,j)}$, $u_{(,j)}$, $v_{(,j)}$ denote the j\textsuperscript{th} column in $A$, $U$, $V$. Let $\tilde{A}$ be the same matrix as $A$ but all $0$'s are replaced by the column mean.

To estimate the parameters in ZIPFA, we propose to maximize the corresponding total zero-inflated Poisson likelihood $L(A)$:
\begin{align}
	L(A)=\prod_{i, j}L(a_{ij}; U, V, \tau, N) &= \prod_{i, j} \left\{ p_{ij}\mathbb{I}(a_{ij}=0)+
	(1-p_{ij})\frac{(N_i \lambda_{ij})^{a_{ij}}e^{-N_i \lambda_{ij}}}{a_{ij}!} \right\} \label{eqn:total_l}
\end{align}
where $\ln (\lambda_{ij})=\sum_{k=1}^K u_{ik}v_{jk}$ and $\operatorname{logit}(p_{ij})=-\tau \ln(\lambda_{ij})$.

In this expression, the scale parameter $\tau$, the factors $U$ and their scores $V$ are all unknown, which makes direct likelihood maximization over all the unknown parameters prohibitive. Hence, we consider an alternating maximum-likelihood algorithm within the generalized linear model (GLM) framework. Specifically, assuming matrix $U$ is known, we transform the optimization problem into a GLM and find the optimal $\tau$ and matrix $V$ that provide maximum $L(A)$;  then we fix matrix $V$ and solve for new $\tau$ and matrix $U$ that maximize $L(A)$ in a similar GLM. This procedure is repeated to increase the total likelihood $L(A)$ until convergence. Since $L(A)$ has a supremum less than $1$, our algorithm is guaranteed to converge. We briefly summarize the model fitting algorithm in the `ZIPFA algorithm' (algorithm 1) box and its details are in Web Appendix A.

\subsection{Zero-inflated Poisson Regression}\label{sec:zero_reg}
In the ZIPFA algorithm, a special type of zero-inflated Poisson regression has been used in step 3 and 4 in the ``ZIPFA algorithm'' (algorithm 1) box above. Now we will present further discussion about this regression. Let the response variable be $Y=(y_1,y_2,\ldots,y_n)^\top$ (i.e., $A^{(v)}$ or $A^{(u)}$ in Section~\ref{sec:max_like}) following a zero-inflated Poisson distribution:
\begin{equation*}
		Y_i\sim \begin{dcases}
	0, & \text{with prob} = p_i\\
	Poisson(m_i \lambda_i), & \text{with prob} = 1-p_i
	\end{dcases}
\end{equation*}
where $m=(m_1,m_2,\cdots,m_n)^\top$ is the known scaling vector (i.e., $N^{(v)}$ or $N^{(u)}$ in Section~\ref{sec:max_like}). Let $X$ be an $n$ by $p$ design matrix without intercept column (i.e., $U^\star$ or $V^\star$ in Section~\ref{sec:max_like}), where column vector $X_i$ denotes the i\textsuperscript{th} row of $X$; $\beta =(b_1, b_2,\ldots, b_{p})^\top$ is  the coefficient vector to be estimated (i.e., $U^s$ or $V^s$ in Section~\ref{sec:max_like}).

With the aforementioned relationship between $p_i$ and $\lambda_i$, the model satisfies:
\begin{equation*}
		\ln \operatorname{E}(Y_i/m_i|X_i)=\ln(\lambda_i)=X_i^\top\beta \qquad \text{and}\qquad 	\text{logit}(p_i)=-\tau\ln(\lambda_i)
\end{equation*}

In  order to estimate parameter $\beta$ and $\tau$, we need to write down the likelihood function to maximize it. A latent variable $Z=(z_1, z_2, \ldots , z_n)$ is introduced to indicate whether $Y_i$ is from true zero or not. Define $z_i=1$ when $Y_i$ is from true $0$; $z_i=0$ when $Y_i$ is from Poisson($\lambda_i$) distribution (i.e., $z_i\sim Bin(1,p_i)$). Then the joint likelihood function of $Y$ and $Z$ is as follows:
\begin{equation*}
	L(Y,Z; \beta,\tau, X) = \prodn p_i^{z_i}\left\{ \frac{(m_i \lambda_i)^{y_i}e^{-m_i \lambda_i}}{y_i!}(1-p_i)\right\} ^{1-z_i}
\end{equation*}

Since we introduce a latent variable $Z$ to the expression, it is natural to exploit an EM algorithm for parameter estimation.

\textbf{E step:} We estimate $z_i$ by its conditional expectation under the current estimates $\beta$ and~$\tau$:
\begin{align*}
	z_i&=\e(z_i;\beta,\tau,X,Y)\notag \\
	& = \begin{dcases}
	 \frac{p_i}{p_i+e^{-m_i\lambda_i}\cdot\ (1-p_i)} &\text{     if } y_i=0\\
		 0 &\text{     if } y_i\ne 0
	\end{dcases}
\end{align*}
When $y_i$ is $0$, the conditional expectation of $z_i$ becomes $\left\{ {1+\operatorname{exp}\left({\tau X_i^\top\beta-m_ie^{X_i^\top\beta}}\right)}\right\}^{-1}$; when $y_i$ is not $0$, we know $p_i$ is $0$ and thus the conditional expectation of $z_i$ equals to $0$.

\textbf{M step:} Now we need to solve the optimal solution to maximize the
conditional expectation of the joint log likelihood function $\ln L(Y,Z, \beta, \tau, X)$ given $Y$, $Z$, $X$.
We apply the Levenberg-Marquardt (LM) algorithm in the optimization, for the reason that this algorithm is quite efficient and more robust than the Newton-Raphson method in many cases
\citep{more1978levenberg}. See Web Appendix B for more technical details.

Finally, we use Frobenius norm of $\beta$ difference between two iterations to indicate convergence and usually an empirical threshold is $1$\textperthousand.\\

\subsection{Rank Estimation}
The number of factors $K$ is selected in a data-driven fashion. When prior knowledge about factor number does not exist, we use cross validation to choose  $K$ in practice
\citep{li2018exponential}.

Suppose the candidate rank set is $\mathbb{K}\subset \mathbb{N}^+$. Let $I_d \in \mathbb{N}^{(nm)}$ be an index set of all elements in $A$ (i.e., $I_d=(11,12,\dots,1m,\cdots,nm)$). Then we randomly divide $I_d$ into $r$ even subsets: $I_{d}^{[1]}$, $I_{d}^{[2]}$ to $I_{d}^{[r]}$. In practice, we usually adopt small $r$ (i.e., $r=5$) to reduce the probability of the situation that a whole row or column is lost in any subsets. If this happens, we will simply redivide $I_d$.

Then we calculate likelihood of the model with rank $\kappa \in \mathbb{K}$ in the $t$\textsuperscript{th} fold~($t\in r$) following the description in the algorithm 2 box. Finally, we sum up the likelihoods of all $r$ folds to obtain the total CV likelihood of the model with rank $\kappa$ and calculate the CV likelihood for every rank $\kappa \in \mathbb{K}$. The number of factors $\kappa$ which provides the maximum CV likelihood is the optimal rank.\\

\begin{minipage}{.98\textwidth}
	\noindent
	\begin{tabularx}{\textwidth}{X}
		\toprule
		\multicolumn{1}{c}{Algorithm 2: The ZIPFA cross validation algorithm in $t$\textsuperscript{th} fold}                    \\
		\hline
		Matrix $A \in \mathbb{R}^{n\times m}$ is to be decomposed to $\kappa$ factors. \\
		\textbf{Initialize:}
		\begin{enumerate}[labelindent=\parindent,leftmargin=*]
			\item   Let $\tilde{A}$ be the same matrix as $A$ but all 0's and elements corresponding to $I_{d}^{[t]}$ are replaced by the column mean of rest values;
			\item  Apply SVD to $\ln(\tilde{A})$ to obtain the components $U^{old}$ and $V^{old}$;
			\item Calculate relative row sum $N$ without elements corresponding to $I_{d}^{[t]}$;
			\item Eliminate the elements with index $I_{d}^{[t]}$ in $A^{(v)}$ (or $A^{(u)}$) and note down their locations. Cross out elements in $N^{(v)}$ (or $N^{(u)}$) on the corresponding location.
		\end{enumerate}
		\textbf{Update:}\\
			
		\hspace{20pt}This part remains the same as regular ZIPFA algorithm described before. \\

		\textbf{CV Likelihood:}
		\begin{enumerate}[labelindent=\parindent,leftmargin=*]
			\setcounter{enumi}{4}
			\item Obtain $\Lambda^{(fit)}=U^{(final)}V^{(final) T}$ and calculate $P^{(fit)}=-\tau \Lambda^{(fit)}$.
			\item Use the distribution assumption in Section~\ref{sec:model_setup} to calculate the likelihood of elements in $A$ with index $I_d^{[t]}$ and sum them up.
		\end{enumerate} \\
		\bottomrule
	\end{tabularx}
\end{minipage}\\

\section{Simulation Study}\label{sec:simulation}
In this section, we illustrate the efficacy of our proposed ZIPFA model through a simulation study. We compare our methods with several other SVD-based methods.

\subsection{Data Generation}
We generate rank-3 synthetic NGS data of $200$ samples ($n=200$) and $100$ taxa ($m=100$) according to the assumption in Section~\ref{sec:model_setup}. The Poisson logarithmic rate matrix $\Lambda=UV^\top$, where $U\in \mathbb{R}^{m\times 3}$ is a left singular vector matrix, and $V\in \mathbb{R}^{n\times 3}$ is a right singular vector matrix. See Web Appendix C for data generation in detail.  Each row in $U$ corresponds to one sample and each row in $V$ indicates one taxon profile.
In Web Figure~2(c) \& (d), we applied complete linkage hierarchical clustering to $U$, $V$
\citep{eisen1998cluster}.
It is clear that both taxa and samples could be clustered into $4$ groups.
For setting (1)--(5), we generate matrix $A^\circ$ that $A_{ij}^\circ \sim Poisson(N_i\lambda_{ij})$ where the scaling parameter $N_i$ is set to be $1$. Also we need  true zero probability $p_{ij}$ to generate inflated zeros.  There are several commonly considered relations between $p_{ij}$ and $\lambda_{ij}$
\citep{lambert1992zero}:

\begin{itemize}[parsep=0pt,itemsep=0pt,topsep=0pt,labelindent=\parindent,leftmargin=*]
	\item \textbf{Setting (1)}. $\operatorname{logit}(p_{ij})=-\tau \ln(\lambda_{ij}) \quad (p_{ij}=\frac{1}{1+\lambda_{ij}^\tau})$
	\item \textbf{Setting (2)}. $\ln\{-\ln(p_{ij})\}=\tau \ln(\lambda_{ij})  \quad (p_{ij}=e^{-\lambda_{ij}^\tau})$
	\item \textbf{Setting (3)}. $\ln\{-\ln(1-p_{ij})\}=\tau \ln(\lambda_{ij})  \quad (p_{ij}=1-e^{-\lambda_{ij}^{-\tau}})$
	\item \textbf{Setting (4)}. $\ln\{-\ln(p_{ij})\}=\ln(\lambda_{ij})  +\ln(\tau) \quad (p_{ij}=e^{-\tau \lambda_{ij}})$.
\end{itemize}

Apart from these four linkages, we examine the performance under the setting where each taxon has a fixed true zero probability $p_j$ that is independent of $\lambda_{ij}$
\citep{b2018glm}:

\begin{itemize}[parsep=0pt,itemsep=0pt,topsep=0pt,labelindent=\parindent,leftmargin=*]
	\item \textbf{Setting (5)}. $p_{j} \sim \text{Unif}(\tau-0.10, \tau+0.10)$.
\end{itemize}

In addition, considering that real biological data are sometimes over-dispersed, we also explore the simulation settings from \citet{b2018glm}'s zero-inflated quasi–Poisson latent factor model. Let $A_{ij}^\circ$ to follow a negative binomial distribution with expectation $N_{i}\lambda_{ij}$ and variance $(N_{i}\lambda_{ij}+N_{i}^2\lambda_{ij}^2\phi_{j})$, where $\phi_{j}$ is the dispersion parameter. The relationship between  true zero probability $p_{ij}$ and Poisson rate $\lambda_{ij}$ is similar to the relationship in setting (4), but it is a positive linkage.
\begin{itemize}[parsep=0pt,itemsep=0pt,topsep=0pt,labelindent=\parindent,leftmargin=*]
	\item \textbf{Setting (6.1)}. $\ln\{ -\ln(p_{ij}) \}=-\ln(\lambda_{ij})+\ln(\tau),\,\phi_j\sim \text{Unif}(0.5, 1.0)$ (low over-dispersion)
	\item \textbf{Setting (6.2)}. $\ln\{ -\ln(p_{ij}) \}=-\ln(\lambda_{ij})+\ln(\tau),\,\phi_j\sim \text{Unif}(1.0, 3.0)$ (high over-dispersion)
\end{itemize}
For all settings above, we adjust the total percentage of excessive zeros by setting different $\tau$ values. Once $p_{ij}$ is generated from these settings, our simulated NGS data matrix $A$ can be obtained by replacing $A^\circ_{ij}$ with $0$ with the probability of $p_{ij}$. Setting (1) is the assumption based on which we develop our model in Section~\ref{method} and all the other settings are misspecified situations.

\subsection{Comparing Methods}\label{compare}
We compare the proposed method with the following SVD-based methods: 1) log-PCA: We first preprocess the data in a typical way: add a small value (e.g., 0.5) to all zeros, and then take a logarithmic value of entries that have been divided by the sum of each row. After that, we apply PCA to the preprocessed matrix; 2) PSVDOS: Poisson singular value decomposition with offset
\citep{lee2013poisson}.
This model is based on regular Poisson factor analysis but automatically 	incorporates sample normalization through the use of offsets; 3) GOMMS: GLM-based ordination method for microbiome samples. This method uses a zero-inflated quasi-Poisson latent factor model and thus is able to handle excessive zeros
\citep{b2018glm}.

\subsection{Simulation Result}
\begin{figure}[!t]
\hspace{-0.08\textwidth}\includegraphics[width=0.55\textwidth]{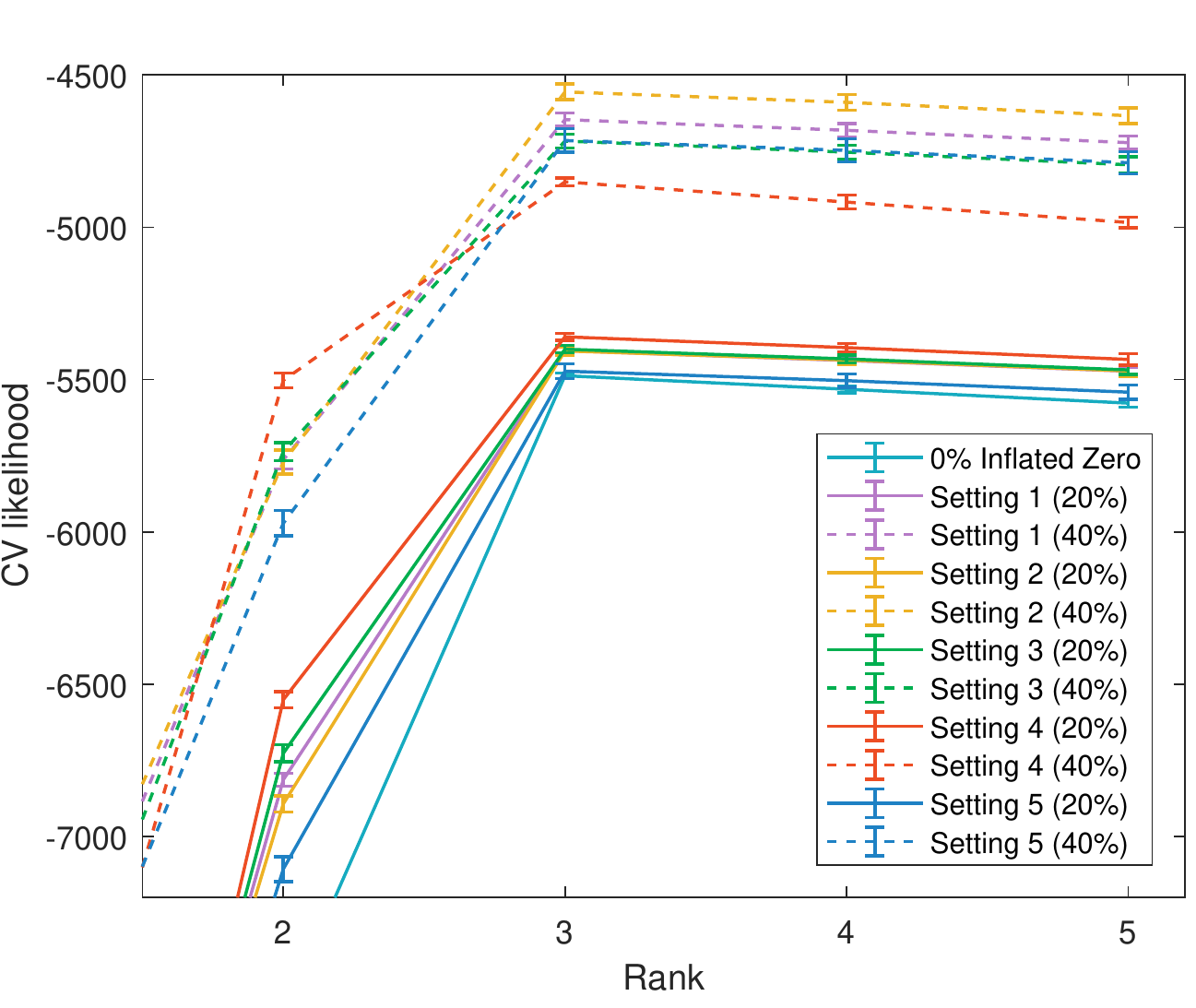}
\includegraphics[width=0.54\textwidth]{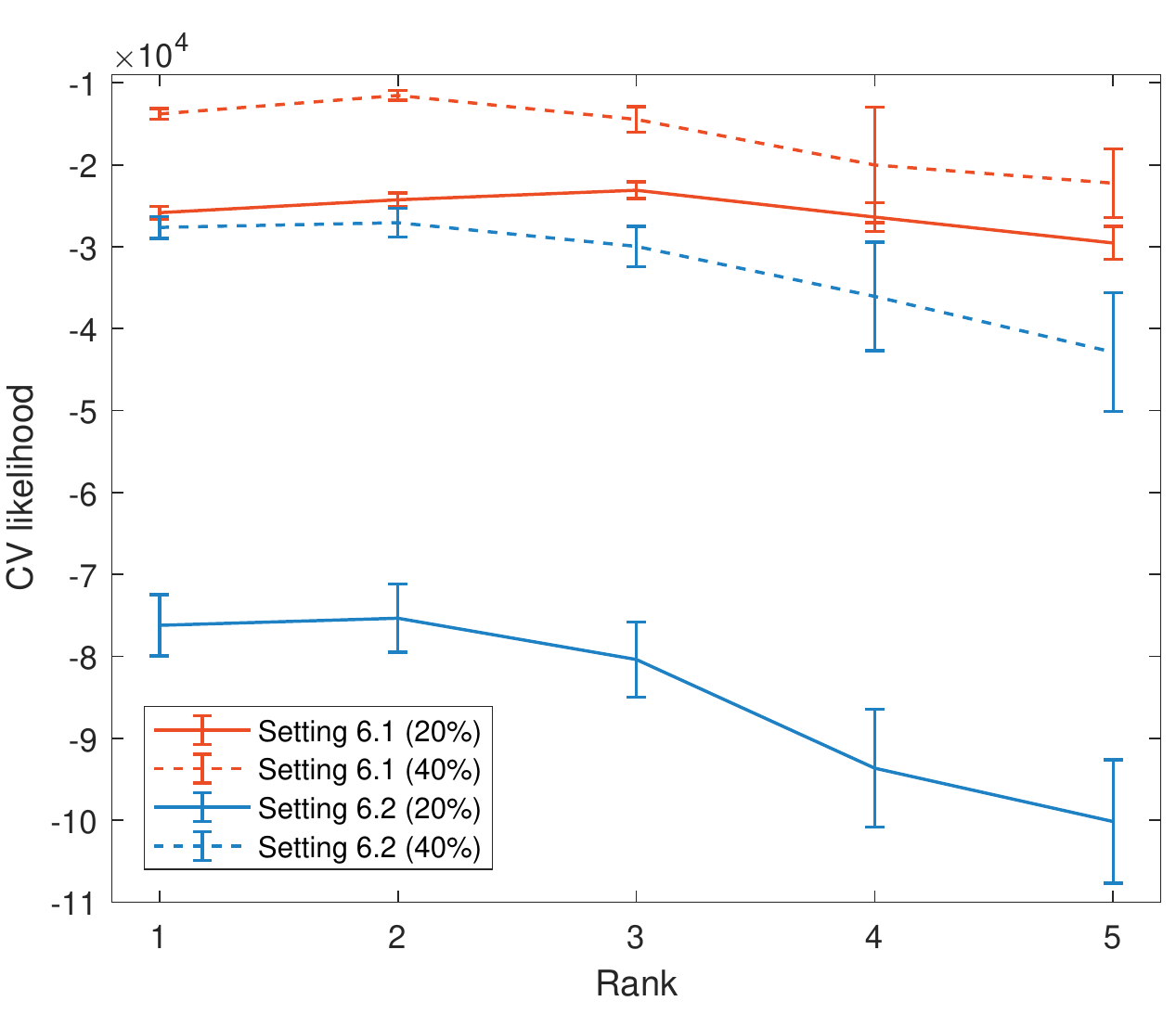}
		\caption{Cross validation to choose the rank in our simulation. Our method provides maximum CV likelihoods with rank $3$ under most simulations settings.}
		\label{fig:cv_simu}
\end{figure}

We first check the rank selection performance of our method. In Figure \ref{fig:cv_simu}, the proposed method provides the maximum CV likelihoods with rank $3$ in most settings except for setting (6.1)($40\%$~inflated zeros) and setting (6.2) where the optimal rank is $2$.  It shows that our method is quite accurate and robust in rank estimation and may underestimate when the model assumption is severely violated.

Then we compare ZIPFA with other models. For all models, we set their ranks to the true rank $3$. GOMMS sometimes has diverged results under some situations, so for each simulation setting, we will conduct enough simulation runs to get at least $200$ converged results. The method performances are evaluated by the Frobenius norm of error matrix and
the clustering accuracy that represents the proportions of taxon/sample that are properly clustered:
\begin{align*}
	\text{L\textsubscript{2} loss} &= \big|\big|\hat U\hat V^\top-\Lambda\big|\big|^2_F \\
	\text{Clustering accuracy } &= \frac{\text{\# of properly clustered taxa/samples}}{\text{\# of total taxa/samples (i.e., 100 taxa/200 samples)}}
\end{align*}
where $\hat U$, $\hat V$ are estimated score and loading matrices; $\Lambda$ is the true natural parameter matrix in the simulation.

In Table~\ref{tab:comparing}, we list the L\textsubscript{2}~loss and the clustering accuracy of taxa and samples under different settings. When  there are no true zeros mixed in the data, all four methods have similar performance and are able to separate the underlying clusters.
Regarding to the L\textsubscript{2}~loss,  in the first four settings, when true zero percent is low (20\%),  all three methods significantly outperform the log-SVD approach, which does not account for the underlying distribution and excessive zeros. When the zero percentage reaches a higher level (40\%),
PSVDOS becomes worse because it could only capture the Poisson part of the data.
ZIPFA has lower or comparable L\textsubscript{2} loss comparing to GOMMS. In setting (5), GOMMS only outperforms ZIPFA when the zero percentage is high (40\%) because this setting essentially favors GOMMS by using independent $\lambda_{ij}$ and $p_{ij}$. In settings (6.1) and (6.2), since the inflated zero probability is positively related with Poisson rate, we obtain negative estimated $\tau$ values in our proposed method. Among four methods, log-SVD is always the worst. When the true zero probability is low (20\%), ZIPFA, PSVDOS and GOMMS have comparable performance and GOMMS has insignificant smaller L\textsubscript{2} loss. When the true zero probability is high, our method outperforms all the other methods.
As for the clustering results, the proposed method is much more appealing than others in the first five settings. In settings (6.1) and (6.2), log-SVD and our method have the best performance, while GOMMS and PSVDOS fail to recover the clustering information in many cases. Overall, our method performs favorably compared to the competing methods even under over-dispersed and/or misspecified simulation settings.

\begin{table}[!p]
	\caption{Comparison of four methods under different settings}
	\label{tab:comparing}
	\centering
	\newcolumntype{Y}{>{\centering\arraybackslash}X}
	\newcolumntype{R}{>{\raggedleft\arraybackslash}X}
	\newcolumntype{L}{>{\raggedright\arraybackslash}X}
		\begin{tabularx}{\textwidth}{crRLRLRLRL}
\toprule
         & Zero \% & \multicolumn{2}{c}{ZIPFA} & \multicolumn{2}{c}{log-SVD} & \multicolumn{2}{c}{PSVDOS} & \multicolumn{2}{c}{GOMMS} \\
\hline
 \multicolumn{10}{c}{L\textsubscript{2} Loss} \\
         & 0\%      & 2.35     & (1.75)   & 2.85     & (0.22)   & \textbf{2.15} & (0.15)   & 4.47     & (0.39) \\
\cmidrule{1-2}\multirow{2}[2]{*}{Setting (1)} & 20\%     & \textbf{4.53} & (0.38)   & 32.84    & (1.56)   & 7.07     & (0.67)   & 6.20     & (0.67) \\
         & 40\%     & \textbf{12.44} & (2.18)   & 171.16   & (4.57)   & 33.93    & (2.57)   & 47.21    & (8.82) \\
\cmidrule{1-2}\multirow{2}[2]{*}{Setting (2)} & 20\%     & \textbf{3.94} & (0.34)   & 37.65    & (1.79)   & 7.65     & (0.78)   & 5.60     & (0.58) \\
         & 40\%     & 28.37    & (6.14)   & 210.88   & (4.42)   & 41.32    & (4.59)   & \textbf{26.74} & (6.87) \\
\cmidrule{1-2}\multirow{2}[2]{*}{Setting (3)} & 20\%     & \textbf{5.10} & (0.41)   & 29.89    & (1.43)   & 6.74     & (0.46)   & 7.25     & (0.77) \\
         & 40\%     & 8.99     & (1.29)   & 144.85   & (3.93)   & 30.35    & (2.46)   & \textbf{8.14} & (0.92) \\
\cmidrule{1-2}\multirow{2}[2]{*}{Setting (4)} & 20\%     & \textbf{7.62} & (1.43)   & 26.56    & (1.26)   & 7.84     & (0.54)   & 10.79    & (1.17) \\
         & 40\%     & \textbf{18.96} & (2.63)   & 95.66    & (2.99)   & 32.34    & (1.49)   & 21.10    & (2.64) \\
\cmidrule{1-2}\multirow{2}[2]{*}{Setting (5)} & 20\%     & \textbf{5.16} & (0.73)   & 46.67    & (3.25)   & 11.26    & (1.91)   & 5.29     & (0.46) \\
         & 40\%     & 18.35    & (3.04)   & 178.06   & (7.09)   & 36.65    & (7.33)   & \textbf{6.23} & (0.61) \\
\cmidrule{1-2}\multirow{2}[2]{*}{Setting (6.1)} & 20\%     & 54.30    & (3.86)   & 158.55   & (4.09)   & 55.91    & (4.16)   & \textbf{49.64} & (16.73) \\
         & 40\%     & \textbf{108.38} & (5.61)   & 425.31   & (4.88)   & 210.84   & (12.82)  & 475.17   & (46.67) \\
\cmidrule{1-2}\multirow{2}[1]{*}{Setting (6.2)} & 20\%     & 52.50    & (3.46)   & 87.89    & (4.29)   & 32.89    & (1.76)   & \textbf{30.50} & (2.08) \\
         & 40\%     & \textbf{111.62} & (5.32)   & 375.38   & (6.31)   & 144.68   & (6.80)   & 256.66   & (39.89) \\
			 \hdashline\\[-1em]
         \multicolumn{10}{c}{Clustering Accuracy by Taxa/Samples} \\
         & 0\%      & \textbf{1.00} & \textbf{1.00} & \textbf{1.00} & \textbf{1.00} & \textbf{1.00} & \textbf{1.00} & \textbf{1.00} & \textbf{1.00} \\
\cmidrule{1-2}\multirow{2}[2]{*}{Setting (1)} & 20\%     & \textbf{1.00} & \textbf{1.00} & \textbf{1.00} & \textbf{1.00} & \textbf{1.00} & \textbf{1.00} & 0.92     & \textbf{1.00} \\
         & 40\%     & \textbf{1.00} & \textbf{1.00} & 0.91     & 0.97     & 0.98     & 0.86     & 0.87     & \textbf{1.00} \\
\cmidrule{1-2}\multirow{2}[2]{*}{Setting (2)} & 20\%     & \textbf{1.00} & \textbf{1.00} & \textbf{1.00} & \textbf{1.00} & \textbf{1.00} & \textbf{1.00} & 0.95     & \textbf{1.00} \\
         & 40\%     & \textbf{1.00} & \textbf{1.00} & 0.66     & 0.72     & 0.92     & 0.77     & 0.92     & \textbf{1.00} \\
\cmidrule{1-2}\multirow{2}[2]{*}{Setting (3)} & 20\%     & \textbf{1.00} & \textbf{1.00} & \textbf{1.00} & \textbf{1.00} & \textbf{1.00} & \textbf{1.00} & 0.91     & \textbf{1.00} \\
         & 40\%     & \textbf{0.99} & \textbf{1.00} & \textbf{0.99} & \textbf{1.00} & \textbf{0.99} & 0.94     & 0.82     & \textbf{1.00} \\
\cmidrule{1-2}\multirow{2}[2]{*}{Setting (4)} & 20\%     & \textbf{1.00} & \textbf{1.00} & \textbf{1.00} & \textbf{1.00} & \textbf{1.00} & \textbf{1.00} & 0.84     & \textbf{1.00} \\
         & 40\%     & 0.96     & \textbf{1.00} & \textbf{1.00} & \textbf{1.00} & \textbf{1.00} & \textbf{1.00} & 0.73     & \textbf{1.00} \\
\cmidrule{1-2}\multirow{2}[2]{*}{Setting (5)} & 20\%     & \textbf{1.00} & \textbf{1.00} & 0.99     & 0.99     & \textbf{1.00} & 0.91     & 0.99     & \textbf{1.00} \\
         & 40\%     & \textbf{1.00} & \textbf{1.00} & 0.77     & 0.83     & 0.94     & 0.82     & 0.93     & \textbf{1.00} \\
\cmidrule{1-2}\multirow{2}[2]{*}{Setting (6.1)} & 20\%     & 0.86     & 0.79     & \textbf{0.94} & \textbf{1.00} & 0.77     & 0.63     & 0.48     & 0.44 \\
         & 40\%     & 0.65     & \textbf{0.93} & \textbf{0.71} & 0.75     & 0.39     & 0.37     & 0.41     & 0.34 \\
\cmidrule{1-2}\multirow{2}[2]{*}{Setting (6.2)} & 20\%     & 0.86     & 0.62     & \textbf{1.00} & \textbf{0.99} & 0.92     & 0.64     & 0.84     & 0.69 \\
         & 40\%     & \textbf{0.77} & 0.73     & 0.69     & \textbf{0.94} & 0.48     & 0.42     & 0.35     & 0.34 \\
			 \bottomrule
	\end{tabularx}
	\raggedright
		\footnotesize
	\hspace{4ex}Note: the best results in each setting are in bold face.
\end{table}

\begin{figure}[!t]
	\centering\includegraphics[width=0.65\textwidth]{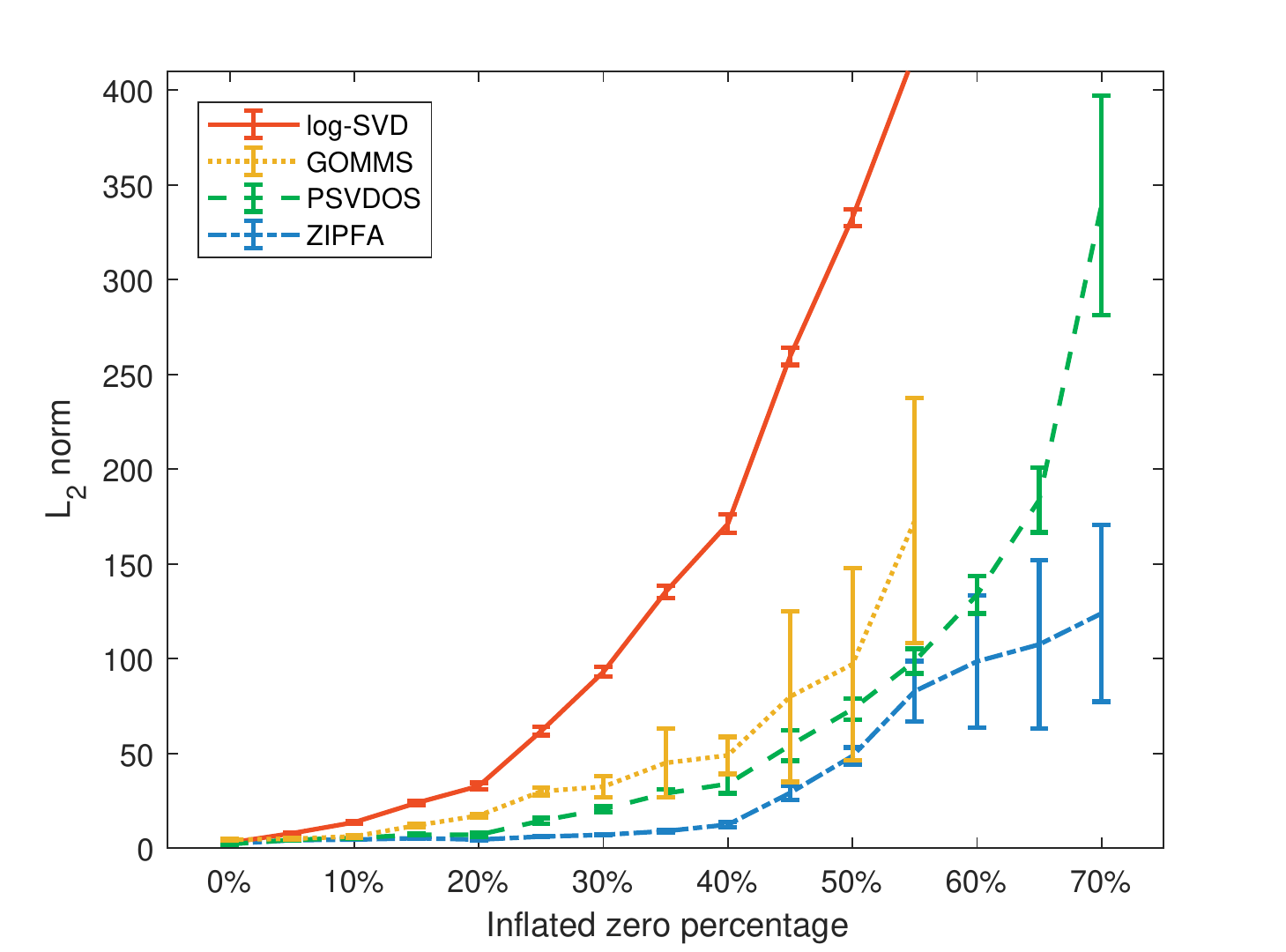}
		\caption{L\textsubscript{2} loss vs. true zero percentage in setting (1). Lower values indicate more accurate fitted results. GOMMS fails to converge when inflated zero percentage exceeds 55\%.}
		\label{fig:zero_proportion}
\end{figure}

Then in Figure~\ref{fig:zero_proportion}, we further explore the performance of the different models for simulated data with different percentages of inflated zeros under setting (1). A typical example of how the fitted results of different models change with increasing inflated zero percentage is shown in Web Appendix D.

In addition, our proposed ZIPFA has favorable convergence property that it successfully converges within moderate iterations in all simulated situations.
GOMMS achieves convergence only for $55\%$, $52\%$, $59\%$ and $71\%$ of the total trials in the first four settings at low zero percentage. When more than half of the data are inflated zeros, we notice that GOMMS fails to converge most of the time (\textgreater$80\%$).

\section{Application to ORIGINS}\label{sec:application}
\subsection{ORIGINS Data}
ORIGINS is a longitudinal cohort study that aims to investigate the cross-sectional association between periodontal microbiota, inflammation and insulin resistance
\citep{demmer2015periodontal}.
In this paper, we will focus on the relationship between subgingival microbial community composition and periodontal disease and identify the bacterial genera associated with some disease indicators.

From February 2011 to May 2013, 300 men and women who met the inclusion criteria were enrolled
\citep{demmer2015periodontal}.
In total, 1,188 subgingival plaque samples (4 samples from 297 participants) were collected from the most posterior tooth per quadrant
and were analyzed using the Human Oral Microbe Identification Microarray to measure the abundances of 379 taxa
\citep{demmer2017subgingival}.
Trained calibrated dental examiners assessed full month attachment loss, probing depth and bleeding on probing at 6 sites per tooth with a UNC-15 manual probe (Hu-Friedy). Other controlled variables include gender, age, ethnicity, education status, BMI and smoking history.

\subsection{Result}
\begin{figure}[!t]
	\parbox{.5\textwidth}{
	\textbf{(a)}\\	
\includegraphics[width=0.5\textwidth]{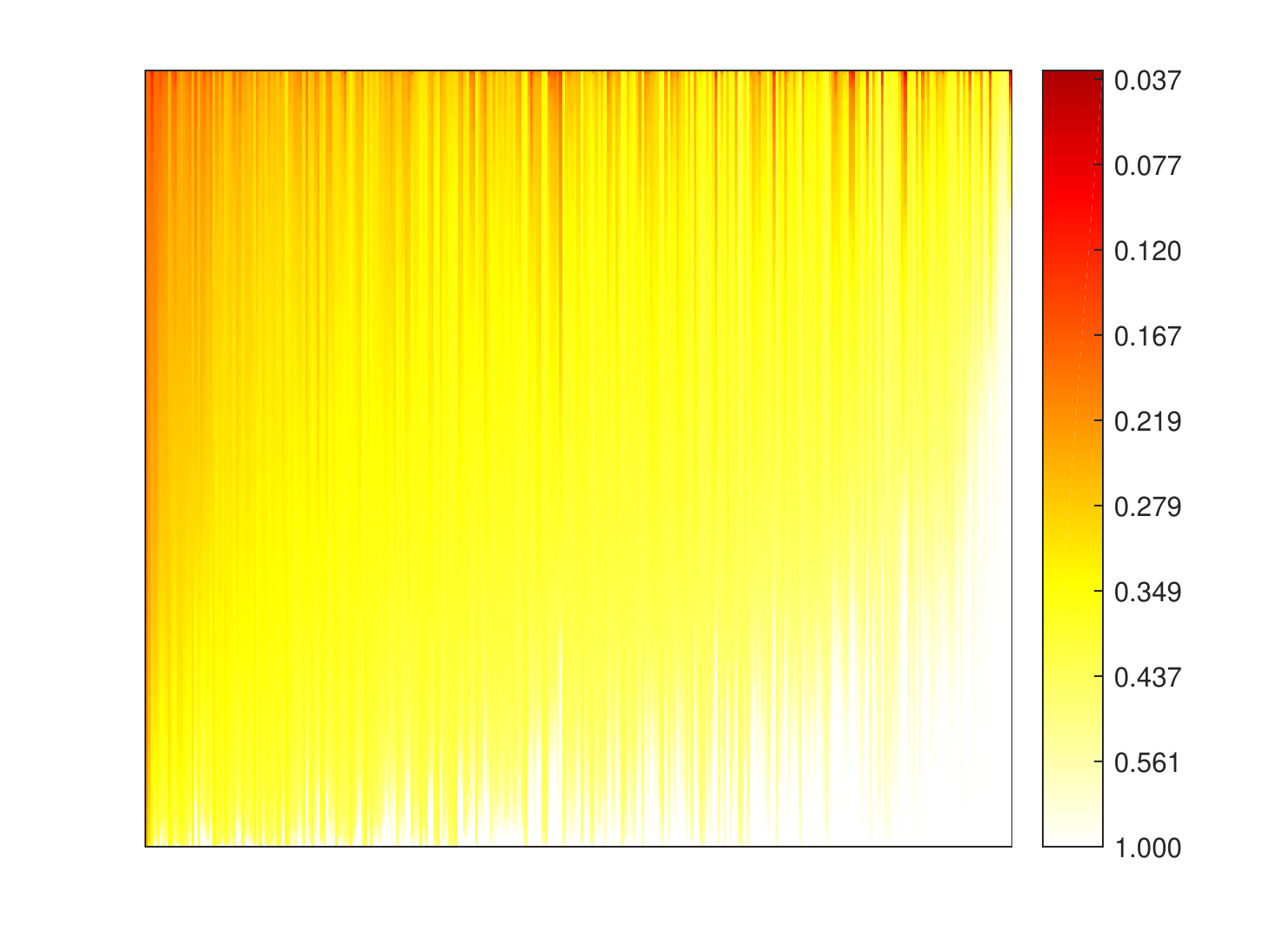}}
\parbox{.5\textwidth}{
	\textbf{(b)}\\
		\includegraphics[width=0.49\textwidth]{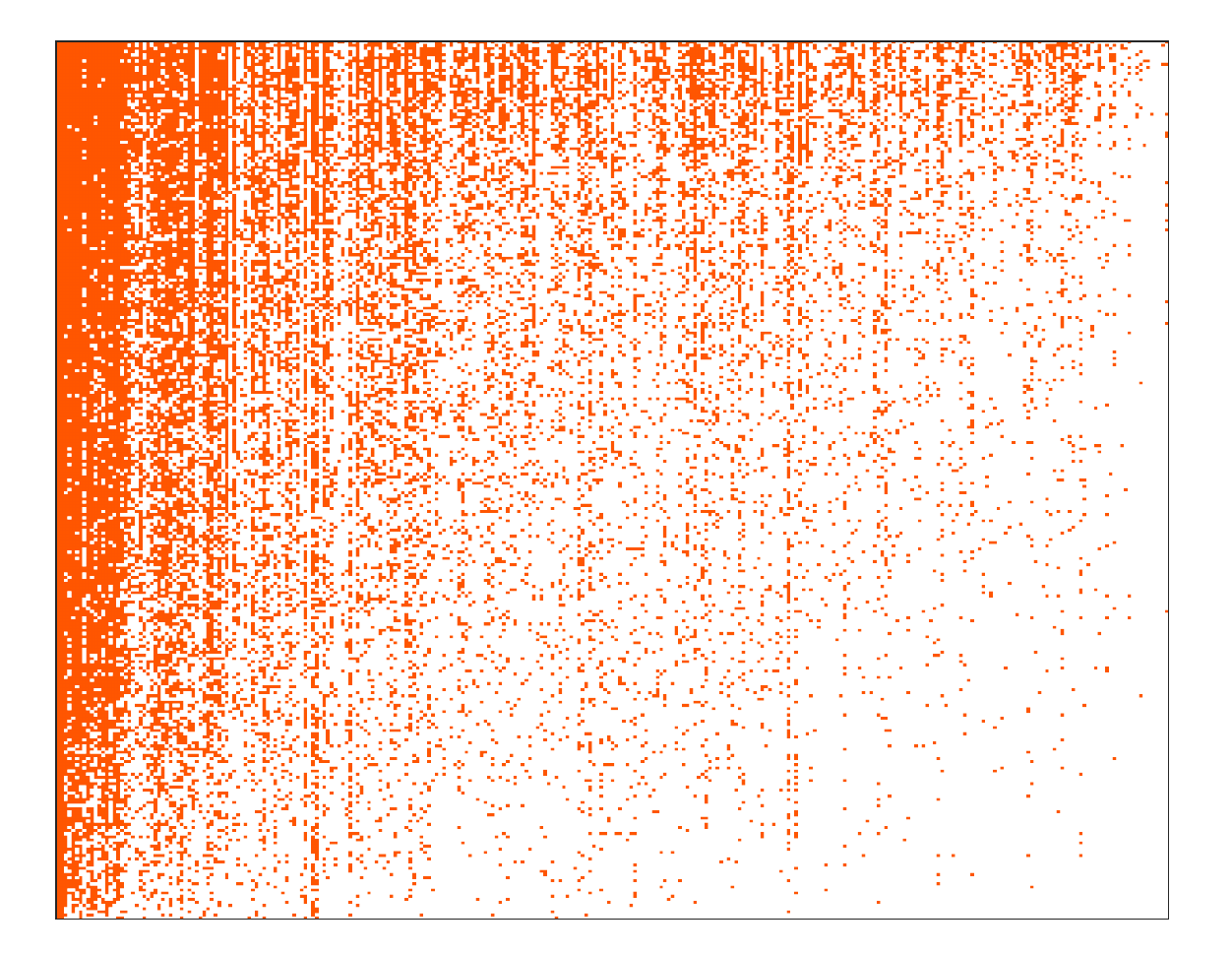}}
\caption{Comparison of predicted  probability  of  zeros  and  real  zero  distribution in the dataset. (a) Heatmap of predicted zero probability. (b) Heatmap of the binary real data value. Red points are non-zero values and white points are zeros. Both heatmaps are rearranged in the same row and column ordering.}\label{fig:comparison}
\end{figure}

We applied 10-fold cross-validation on the data. Web Figure 4 shows that CV likelihood reaches the maximum point at rank equal to 5, so we will use five factors in the following analysis.
We fit a rank-5 ZIPFA to the absolute microbiome data. The algorithm converges after 7 iterations (likelihood change $<1$\textperthousand) in 30 seconds (Matlab R2017a, i9-7900X with 32GB memory). The proposed model gives us the estimated score and loading matrix. With such information, we are able to recover the $\Lambda$, $P$ matrix according to the model assumption in Section~\ref{sec:model_setup}. The total estimated probability of being zero for each count is $\hat p_{ij}+e^{-\hat \lambda_{ij}}$. We reorder the larger values in total zero probability matrix to top-left, and put the smaller values to bottom-right. The heatmap of reordered total zero probability is plotted in Figure~\ref{fig:comparison}(a) and the true data with the same rearrangement is in Figure~\ref{fig:comparison}(b). We compare the predicted probability of zeros with real data zero distribution to examine the level of similarity. A good resemblance indicates that our methods well captures the structure of the excessive zeros.

In order to find the association between five factors obtained by ZIPFA and three responses (full mouth mean attachment loss, mean probing depth and bleeding on probing), we fit linear models. In each model, a response variable is regressed on all five factors and six additional covariates including gender, age, ethnicity, education status, BMI and smoking history. In Web Table 1, p-values corresponding to different factors and response variables are listed. As a comparison, we also demonstrate the result of other methods that we introduce in Section \ref{compare} including log-SVD, PSVDOS, GOMMS and two widely used traditional methods including principal coordinates analysis (PCoA) and non-metric multidimensional scaling (nMDS) on Bray-Curtis  distance \citep{doi:10.2307/1942268}. NMDS in this case fits the data quite well with stress score 0.098 (Shepard stress plot is in Web~Figure~5). The major drawback of PCoA and nMDS is that they are only useful for dimension reduction and data visualization, but cannot identify the explicit relations between the reduced factors and the original taxa. In other words, they cannot provide ``loadings" as in a factor model. We observe ZIPFA, log-SVD, PCoA and nMDS provide significantly associated factors with each response while PSVDOS and GOMMS fail to find a significant factor for ``full mouth mean attachment loss''. In particular, factor 2, 3 in the our proposed method are significant predictors of all these periodontal disease indicators, which may imply a potential link in oral microbiome composition and periodontal disease.

\begin{figure}[!t]
	\textbf{(a)}\\
	\includegraphics[width=1\textwidth,trim=1cm .0cm 1.8cm 0.5cm,clip]{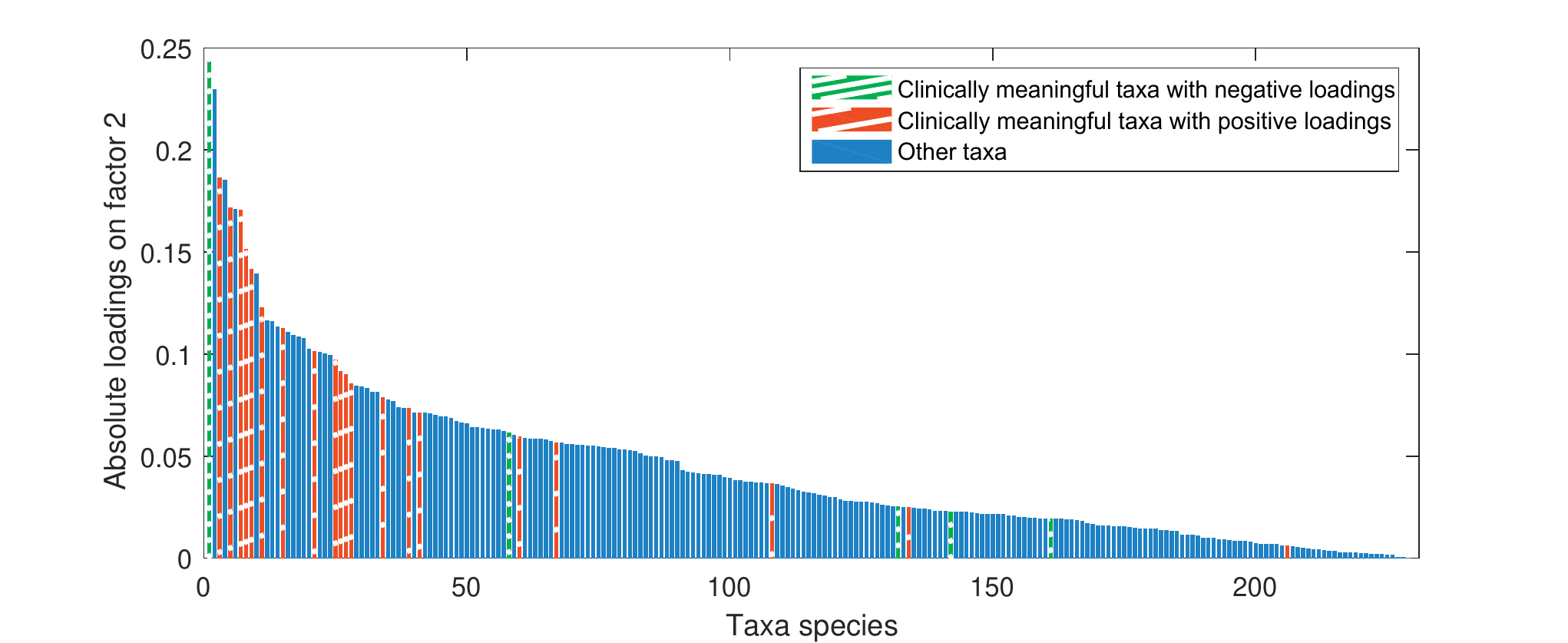}
	\textbf{(b)}\\
	\includegraphics[width=1\textwidth,trim=1cm .0cm 1.8cm .5cm,clip]{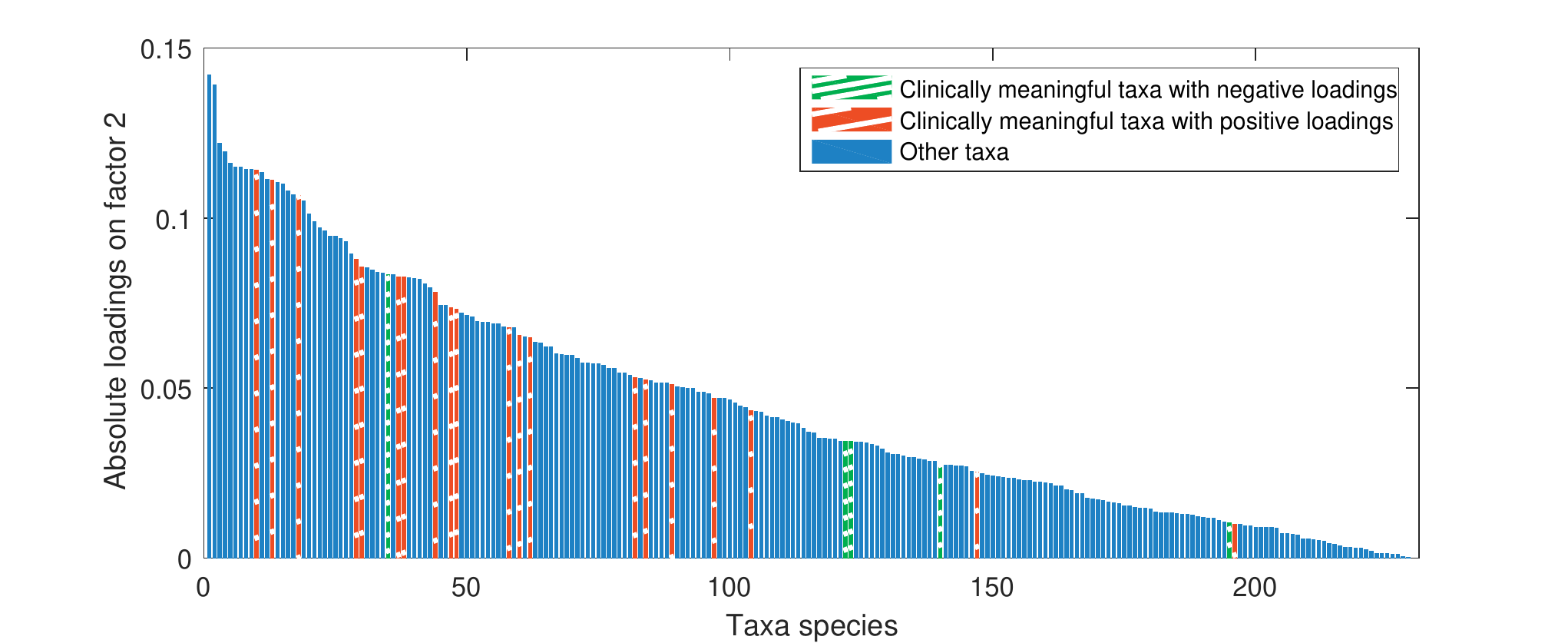}
		\caption{Absolute taxa loadings on the most significant factor. Each bar is a loading value of the factor. Blue or red bars are clinically meaningful taxa in published literature. (a) Loadings on factor 2 of our proposed ZIPFA. (b) Loadings on factor 2 of log-SVD.}
		\label{fig:taxa}
\end{figure}

Due to the nature of distance-based ordination methods, PCoA and nMDS do not help to select taxa that are potentially relevant to periodontal diseases. For the rest of the methods, we further look into the loading vector corresponding to the significant factors to identify important taxa related to periodontal disease.
In previous literature, many researchers have studied the associations between subgingival microbiota and periodontal disease.
\citet{wade1996role}
showed that asaccharolytic species \textit{E. brachy}, \textit{E. nodatum}, \textit{E. saphenum} and \textit{E. timidum} have significant associations with severe periodontal disease, but are only rarely found in healthy oral sites.
\citet{socransky1998microbial}
pointed out  a group of taxa called ``red complex": \textit{P. gingivalis}
, \textit{T. denticola} and \textit{T. forsythia}
are frequently isolated together and are strongly associated with diseased sites in the mouth.
\citet{lombardo2012porphyromonas}
discovered the presence of  \textit{P. endodontalis} is associated with chronic periodontitis.
\citet{dani2016assessment}
found that \textit{S. mutans} is considered to be one of the major pathogens in dental caries and is very likely to induce periodontal disease.  A molecular survey by Griffen revealed new species \textit{Filifactor alocis} is strongly associated with periodontal disease
\citep{griffen2012distinct}.
Black-pigmented gram-negative anaerobes including \textit{P. asaccharolytica}, \textit{P. endodontalis}, \textit{P. gingivalis}, \textit{P. buccae}, \textit{P. intermedia}, \textit{P. melalinogenica}, and \textit{P. nigrescens} can be easily seen in adults with rapidly progressing periodontitis lesions and are correlated with sheep periodontitis
\citep{borsanelli2017black}.
\citet{kumar2003new}
stated that associations with chronic periodontitis were observed for several new species including \textit{P. denticola}, \textit{C. curtum}, \textit{A. rimae} and \textit{A. parvulum}. In addition,
\citet{abusleme2013subgingival}
tested a group of taxa and we take the most significant 10 taxa as clinically interesting species. We plot the absolute loadings of all taxa on the most significant associated factor of ZIPFA and log-SVD in Figure~\ref{fig:taxa} and on the two most significant factors of all four methods in Web Figure 6. The taxa that are potentially clinically meaningful are marked in red (positive association) and green (negative association). Conceptually, those taxa should have large absolute loading values corresponding to the significant factors. In Figure~\ref{fig:taxa},  clearly, for ZIPFA, clinically meaningful taxa tend to concentrate on the left with large loading values while for log-SVD, they are more scattered (See Web Table 2 for more details). A permutation test of the mean ranks of the relevant taxa further shows that the difference is significant (p-value $=1.57\times 10^{-5}$). Namely, the result from our method is more consistent with the literature. Similarly, In Web Figure~6, we see that ZIPFA separates the clinically meaningful taxa with larger absolute loadings. Log-SVD fails to pick out most negative associated taxa, and rest of the methods have inferior results as well.
Our method also suggests that further investigation into the following taxa is justified: \textit{Lachnoanaerobaculum sp. HOT 083}, \textit{Leptotrichia sp. HOT 219}, \textit{N. pharyngis} and \textit{Bacteroidetes[G-5] sp. HOT 511}.

\section{Discussion}\label{sec:dicussion}
Dimension reduction is a common feature of many microbiome analytical workflows
\citep{cao2017microbial}.
This paper presents an new method of factor analysis which takes the distribution of counts into full consideration. The proposed model includes one shape parameter~($\tau$) to link the true zero probability and Poisson expectation and achieves satisfactory fitting on the data.
In addition, the zero-inflated Poisson regression proposed in Section~\ref{sec:zero_reg} is a new method in zero-inflated regression analysis.
We also develop a new CV approach for estimating the rank of the underlying natural parameter matrix. In the ORIGINS analysis, the proposed method identifies microbial profiles that are significantly associated with clinical outcomes and generates new scientific hypotheses for lab research.

There are several future research directions worth studying.  While the method is developed for count data in microbiome studies, this idea can be extended to other situations. For example, a zero-inflated negative binomial  distribution can be considered when the data has extra dispersion
\citep{srivastava2010two}.
We can also change the logistic link function to others if the relationship between $\operatorname{logit}(p_{ij})$ and $\ln(\lambda_{ij})$ is not linear. But how to choose the best link is still a remaining question. In addition, we can reduce the computation cost of ZIPFA by further optimizing the EM algorithm in the future.

\newpage

\section*{Reference}
Reference is at the end of this document. 

\section*{Acknowledgements}
Research reported in this publication was supported by the National Institute Of Dental \& Craniofacial Research of the National Institutes of Health under Award Number R03DE027773.

\section*{Supporting Information}
Additional supporting information may be found online in the Supporting Information section at the end of the article, including Web Appendices, Tables and Figures referenced in Sections 2-4. The Matlab codes in the paper and some examples can be found on Github: \url{https://github.com/zjph602xtc/ZIPFA}. The R package link: \url{https://cran.r-project.org/web/packages/ZIPFA/index.html}.

\renewcommand*\footnoterule{}
\renewcommand{\figurename}{Web Figure.}
\renewcommand{\tablename}{Web Table.}
\graphicspath{{./paper/}{./paper_new/}}

\title{\bf Web-based Supplementary Materials for ``Zero-inflated Poisson Factor Model with Application to Microbiome Absolute Abundance Data"}

\author{Tianchen Xu$^{*}$\\
Mailman School of Public Health, Columbia University, NY 10032, USA
\and 
Ryan T. Demmer\\
School of Public Health, University of Minnesota, MN 55454, USA
\and
Gen Li\\
Mailman School of Public Health, Columbia University, NY 10032, USA
}

\maketitle
\setlength{\abovedisplayskip}{3pt}
\setlength{\belowdisplayskip}{3pt}
\setlength{\abovedisplayshortskip}{3pt}
\setlength{\belowdisplayshortskip}{3pt} 

\newpage
\setcounter{section}{0}
\section{Web Appendix A}
In order to obtain the initial $U$ and $V$, we apply the singular value decomposition (SVD) to the log-transformed matrix $\tilde{A}$ and obtain the components $\{U', V', S'\}$ (i.e., $\ln(\tilde{A})= U'S'V'^\top$). Set $V^{old}=V'$ and $U^{old}=(s'_{11}u'_{(,1)}, s'_{22}u'_{(,2)},\ldots, s'_{KK}u'_{(,K)})$, where $s'_{kk}$ is the $k$-th diagonal element of $S'$. Assuming matrix $U^{old}$ is known, we estimate the loadings $v_{(j,)}$ and shape parameter $\tau$ by fitting a ZIP regression (will be discussed in Section~2.3) with $a_{(,j)}$ as the response, vector $u^{old}_{(,1)}$, $u^{old}_{(,2)}$, $\cdots$, $u^{old}_{(,K)}$ as the covariates and a scaling vector $N$ as an extra offset parameter. Since $A$ has $m$ columns, we need to fit $m$ GLMs to obtain $m$ rows in $V$. 
However, an important assumption of our model is that the link between $p_{ij}$ and $\lambda_{ij}$ (i.e., $\tau$) remains the same across all $m$ different GLMs. To accommodate this, we solve all $m$ models simultaneously to get a globally best $\tau$ value.
Here we combine the response, covariates and offset parameter in $m$ regressions into larger scale matrices $A^{(u)}$, $U^{\star\,old}$ and $N^{(u)}$ and fit a ZIP regression with $A^{(u)}$ as response variable, columns in $U^{\star\,old}$ as covariates and  $N^{(u)}$  as the offset parameter:
\begin{equation*}
	A^{(u)}= \begin{pmatrix}
	a_{(,1)}\\
	a_{(,2)}\\
	\vdots\\
	a_{(,m)}
	\end{pmatrix}, \qquad
	U^{\star\,old}= \underbrace{ \begin{pmatrix}
	U^{old} &&&\\
	&\hspace{-10pt}U^{old} &&\\
	&&\hspace{-15pt}\ddots &\\
	&&\hspace{-10pt}&U^{old}
	\end{pmatrix}}_{m \text{ times}}, \qquad
	N^{(u)}=\left. \begin{pmatrix}
		N\\
		N\\
		\vdots\\
		N
	\end{pmatrix}\right \} {\scriptstyle m \text{ times  }}.
\end{equation*}
The fitted coefficient $V^s$ is a combined vector of $v_{(i,)}$ such that $V^s=(v_{(1,)}, v_{(2,)}, \cdots, v_{(m,)})$. We could obtain the fitted $V^{new}$ by cutting $V^s$ to $v_{(i,)}$'s and rearrange $v_{(i,)}$'s to matrix $V^{new}$. 

Then update $U^{new}$ in a similar fashion. The response and covariates are $A^{(v)}$ and every columns in $V^{\star\,new}$ and an offset parameter $N^{(v)}$ is needed as well:
\begin{equation*}
		A^{(v)} = \begin{pmatrix*}
	a_{(1,)}^\top\\[5pt]
	a_{(2,)}^\top\\[5pt]
	\vdots\\[5pt]
	a_{(n,)}^\top
	\end{pmatrix*}, \quad
	V^{\star\,new} =\underbrace{ \begin{pmatrix}
	V^{new} &&&\\
	&\hspace{-10pt}V^{new} &&\\
	&&\hspace{-20pt}\ddots &\\
	&&&\hspace{-10pt}V^{new}
	\end{pmatrix}}_{n \text{ times}}, \quad
	N^{(v)} = \bigg(
	\underbrace{N_1,
	\cdots
	N_1}_{m \text{ times}},
	\cdots,
	\underbrace{N_n,
	\cdots,
	N_n}_{m \text{ times}}
\bigg)^\top.
\end{equation*}
The fitted coefficient is $U^s=(u_{(1,)}, u_{(2,)}, \cdots, u_{(n,)})$ and thus the  $U^{new}$ is able to be reconstructed from $U^s$ in a similar way like $V^{new}$. 


After $U$, $V$ are updated, one more step is involved to ensure the uniqueness and orthogonality of these updated components. We apply SVD to the $U^{new}V^{new\,T}$ and label the components by $\{U', V', S'\}$. Set $U^{old}=(s'_{11}u'_{(,1)}, s'_{22}u'_{(,2)},\ldots, s'_{KK}u'_{(,K)})$ and $V^{old}=V'$. 

We use the updated $U$, $V$ and $\tau$ to obtain the estimates of $\Lambda$ and $P$ in the current round of iteration and then calculate the likelihood value $L(A)$: 
\begin{align*}
	L(A)=\prod_{i, j}L(a_{ij}; U, V, \tau, N) &= \prod_{i, j} \left\{ p_{ij}\mathbb{I}(a_{ij}=0)+
	(1-p_{ij})\frac{(N_i \lambda_{ij})^{a_{ij}}e^{-N_i \lambda_{ij}}}{a_{ij}!} \right\} \label{eqn:total_l}
\end{align*}
where $\ln (\lambda_{ij})=\sum_{k=1}^K u_{ik}v_{jk}$ and $\operatorname{logit}(p_{ij})=-\tau \ln(\lambda_{ij})$. 

When the percentage of total likelihood difference between two iterations is less than a certain small value, the algorithm terminates; Otherwise, we continue to update $U$, $V$, $\tau$ until convergence.  In ZIP regression step where $U$, $V$ and $\tau$ are updated, we will use the EM algorithm to estimate the coefficients (see Section~2.3), and thus the likelihood increases due to the nature of EM algorithm used in regression estimation. 
The likelihood remains the same in SVD step. Overall, the algorithm is guaranteed  to converge.

\section{Web Appendix B}
LM algorithm introduces a positive damping parameter $\mu$. If we reduce $\mu$, the LM algorithm behaves like Newton's method, which is a good way to get quadratic convergence in the final 
stages of the iteration; while if we enlarge $\mu$, the descent direction in LM algorithm is closer to the gradient descent method, which is free from the information of second derivative of the objective function and thus the algorithm still works when the Hessian matrix of the objective function is ill-conditioned or nearly singular.  Such an algorithm could be put into the trust-region framework and is implemented in many solvers with common programming languages
\citep{yuan1999nonlinear, more1983computing}. \\

\noindent\textbf{B.1 LM Algorithm}\\
We use Levenberg-Marquardt method to solve $\beta$ and $\tau$. 
The objective function is $Q(\beta,\tau)=-\ln(L)$ and we want to minimize the objective function. The log likelihood function is:
	\begin{align*}
\ln(L) & = \sumn z_i \ln(p_i) + \sumn (1-z_i)\big\{ y_i\ln(m_i \lambda_i)-m_i\lambda_i-\ln(y_i!) + \ln(1-p_i) \big\} \notag \\
	&= -\tau Z X\beta - \sumn \ln(1+e^{-\tau X_i^\top \beta})\notag \\
	& \qquad +(1-Z)\{\text{diag}(y_i)(X\beta+\ln(m))-\text{diag}(m_i)e^{X\beta}-\ln(Y!)\} 
\end{align*}
where $\operatorname{diag(\alpha_i)}$ represents a matrix whose diagonal elements are $\alpha_i$; $\ln$ is a pointwise operator on the vector.

The  first and second derivative of $\ln(L)$ are:	
\begin{align*}
	J_\beta =\frac{\partial \ln(L)}{\partial \beta^\top} &= \left(-\tau Z+\tau W +(1-Z)U \right)X\\
	J_\tau = \frac{\partial \ln(L)}{\partial \tau} &= (W-Z)X\beta\\
	H_{\beta\beta}	=\frac{\partial^2 \ln(L)}{\partial \beta \partial \beta ^\top}&= X^\top R X\\
	H_{\tau\tau}=\frac{\partial^2 \ln(L)}{\partial\tau^2}&=-\sumn\frac{(X_i^\top\beta)^2e^{\tau X_i^\top\beta}}{(e^{\tau X_i^\top\beta}+1)^2}\\
	H_{\tau\beta}=\frac{\partial^2 \ln(L)}{\partial \tau \partial \beta^\top}&=(V-Z)X
\end{align*}
where $W=\big((e^{\tau X_1\beta}+1)^{-1},\ldots,(e^{\tau X_n\beta}+1)^{-1}\big)$,\\
\phantom{where }$U=\text{diag}\left(y_i-m_ie^{X_i^\top \beta}\right)$,\\
\phantom{where }$R=\text{diag}\Big(-\frac{\tau^2e^{\tau X_i^\top\beta}}{(e^{\tau X_i^\top\beta}+1)^2}-(1-z_i)(m_ie^{X_i^\top\beta})\Big)$,\\[5pt]
\phantom{where }$V=\Big(\frac{e^{\tau X_1 \beta}-\tau X_1\beta e^{\tau X_1\beta}+1}{(e^{\tau X_1 \beta}+1)^2},\ldots,\frac{e^{\tau X_n \beta}-\tau X_n\beta e^{\tau X_n\beta}+1}{(e^{\tau X_n \beta}+1)^2}\Big)$. \\
In addition, $\operatorname{diag(\alpha_i)}$ represents a matrix whose diagonal elements are $\alpha_i$; $\ln$ is a pointwise operator on the vector.

Correspondingly, the Jacobian (J) and Hessian (H) matrix of objective function $Q(\beta,\tau)$are:
\begin{align*}
	J(\beta,\tau) = -\begin{pmatrix}
		J_\beta^\top\\ J_\tau
	\end{pmatrix}  \qquad 
	H(\beta,\tau) = -\begin{pmatrix}
		H_{\beta\beta} & H_{\tau\beta}^\top\\
		H_{\beta\tau} & H_{\tau\tau}
	\end{pmatrix}
\end{align*}
\textit{Initialize:}\\
\indent \textit{Step 1:} Set initial $\beta_0$ and $\tau_0$, which is discussed in Section~2.3. Then we calculate $J_0$ with these initial values. \\
\indent \textit{Step 2:} Set initial damping parameter $\mu$:
\begin{align*}
	\mu= \rho \cdot J_m
\end{align*}
where $\rho$ is user specified and the default value is $1\times 10^{-5}$; $J_m$ is the maximum element in matrix $J_0^\top J_0$.\\
\textit{Iterating:}\\
\indent \textit{Step 3:} Multiply $\mu$ by $2$ until $(H(\beta,\tau)+\mu I)$ is positive definite.\\
\indent \textit{Step 4:} Obtain the descent direction $h$ by solving equation:
\begin{align*}
	(H(\beta,\tau)+\mu I) h=-J(\beta,\tau).
\end{align*}
\indent \textit{Step 5:} Define and compute the gain ratio $\delta$:
\begin{align*}
	\delta = -\frac{Q(\beta,\tau)-Q((\beta,\tau)+h)}{h^\top J(\beta,\tau)+\frac{1}{2}h^\top(H(\beta,\tau)+\mu I)h}.
\end{align*}
\indent \textit{Step 6:} 
If $\delta > 0.001$, we obtain better $\ln L(\beta,\tau)$ value, so we update $\beta, \tau$, $\mu$ by:
\begin{align*}
	\begin{pmatrix}
		\beta_{k+1} \\ \tau_{k+1}
	\end{pmatrix} = \begin{pmatrix}
		\beta_{k}\\ \tau_{k}
	\end{pmatrix} + h, \qquad \mu = \mu \cdot \max{\left\{\frac13, 1-(2\delta-1)^3\right\} }
\end{align*}
\phantom{\indent \textit{Step 6:}}If $\delta \le 0.001$, do not update $\beta, \tau$. Update $\mu = 2\mu$.\\
\indent \textit{Step 7:} Repeat from step 3 to step 6 until the algorithm converges. The convergence criterion is discussed in Section~2.3.\\

\noindent\textbf{B.2 Initial Values}\\
Similar to many fitting algorithms, the LM algorithm only finds a local extremum, which requires us to provide a reasonable initial $\tau_0$ and $\beta_0$. Fitted $\beta$ from Poisson regression could be used as the initial value in the first iteration of the EM algorithm. For initial value of $\tau_0$, we have to estimate $\bar p$ first. Assuming that all $y_i$ from the Poisson distribution are not zero, estimate $\bar p$ under such an assumption: 
\begin{equation*}
		\bar p=\frac{\text{\# }y_i\text{'s that are zeros}}{n}
\end{equation*}
Obviously, this expression will overestimate the $\bar p$, because zeros from the Poisson distribution are wrongly regraded and counted as a part of true zeros, especially when the estimated $\bar p$ is relatively low. However, such a estimator is usually accurate enough to help the fitting algorithm to converge. Then we are able to estimate the initial $\tau_0$ at beginning of each round by the following expression (the derivation is at end of Web Appendix B):
\begin{equation}\label{eq}
	\tau_0=-\frac{n \operatorname{logit}(\bar p)} {\sumn X_i^\top\beta}
\end{equation}
In practice, if the model is stable and the algorithm converges successfully with all initial parameters ($\tau$ and $\beta$) in the previous iteration, we tend to use the fitted the parameters in the previous iteration, for which will accelerate the convergency speed. Only if using the previous parameters as the initial parameter causes divergence, we will then turn to the estimation of $\beta_0$ and $\tau_0$ mentioned above.\\

\noindent\textit{Derivation of equation~\eqref{eq}:}\\
Since we have the relationship between $p_i$ and $\lambda_i$:
\begin{align*}
	\operatorname{logit}(p_i)=-\tau \ln(\lambda_i)
\end{align*}
Plug $\ln(\lambda_i)=X_i^\top \beta$ into the expression above:
\begin{align*}
	\operatorname{logit}(p_i)=-\tau X_i^\top \beta
\end{align*}
We use $\operatorname{logit}(\bar p)$ to present the mean value on the left hand side, and take the average of $X_i\beta$:
\begin{align*}
	\operatorname{logit}(\bar p)=-\frac{\tau}{n} \sum_{i=1}^n X_i^\top \beta
\end{align*}
Correspondingly, the estimation of $\tau_0$ is:
\begin{equation*}
	\tau_0=-\frac{n \operatorname{logit}(\bar p)} {\sumn X_i^\top\beta}
\end{equation*}

\section{Web Appendix C}
We generate rank-3 synthetic NGS data of $200$ samples ($n=200$) and $100$ taxa ($m=100$) according to the assumption in our paper. The Poisson logarithmic rate matrix $\Lambda=UV^\top$, where $U\in \mathbb{R}^{m\times 3}$ is a left singular vector matrix, and $V\in \mathbb{R}^{n\times 3}$ is a right singular vector matrix. We consider three different clustering patterns in the samples as depicted in $U$. To generate $U$, we create a 200-by-3 matrix $U$ such that:
\begin{alignat*}{2}
	&U(36:80,1)=2.0,&\qquad &U(81:140,1)=1.7\\
	&U(1:35,2)=1.8,&\qquad &U(36:80,2)=0.9\\
	&U(36:200,3)=1.7&&
\end{alignat*}
with all the other entries being 0, and then jitter all the entries
by adding random numbers generated from $N(0, 0.06^2$). Similarly, To generate $V$, we create a 100-by-3 matrix $V$ such that:
\begin{alignat*}{2}
	&V(61:100,1)=1.7& &\\
	&V(36:60,2)=1.7,&\qquad &V(61:100,2)=1.0\\
	&V(1:25,3)=1.7,&\qquad& V(26:100,3)=0.9
\end{alignat*}
with all the other entries being 0, and then jitter all the entries
by adding random numbers generated from $N(0, 0.05^2)$. The three columns of $U$ and $V$ are plotted in the columns of Web Figure~\ref{u_v}(a) and the true $\ln(\lambda)$ matrix is plotted in Web Figure~\ref{u_v}(b). Each row in $U$ corresponds to one sample and each row in $V$ indicates one taxon profile. 
In Web Figure~\ref{u_v}(c) \& (d), we applied complete linkage hierarchical clustering to $U$, $V$
\citep{eisen1998cluster}. 
It is clear that both taxa and samples could be clustered into $4$ groups. 

\section{Web Appendix D}
As an example, Web Figure \ref{fig:different_method} shows a typical case that how the fitted distribution changes in setting (1) as inflated zeros percentages go from 0\% to 40\%. All methods work well when inflated zeros do not exist. When the true zero percentage goes higher (20\%), the estimated distribution from log-SVD shifts to the left to capture the excessive zeros. When true zero percentage continues growing to 40\%, our method is the only method that could keep the right clustering of both samples and taxa. log-SVD fails to recover the underlying Poisson rate distribution under such a high true zero percentage. PSVDOS performs
better to some extent, but also fails to capture the right taxa clustering just like log-SVD.
GOMMS successfully clusters the taxa, but the sample clustering is not reflected because
it assigns each taxon a single probability of true zero and this might limit the heterogeneity
between samples.

\setcounter{figure}{0}
\setcounter{table}{0}
\begin{figure}[!p]
	\centering\includegraphics[width=0.75\textwidth]{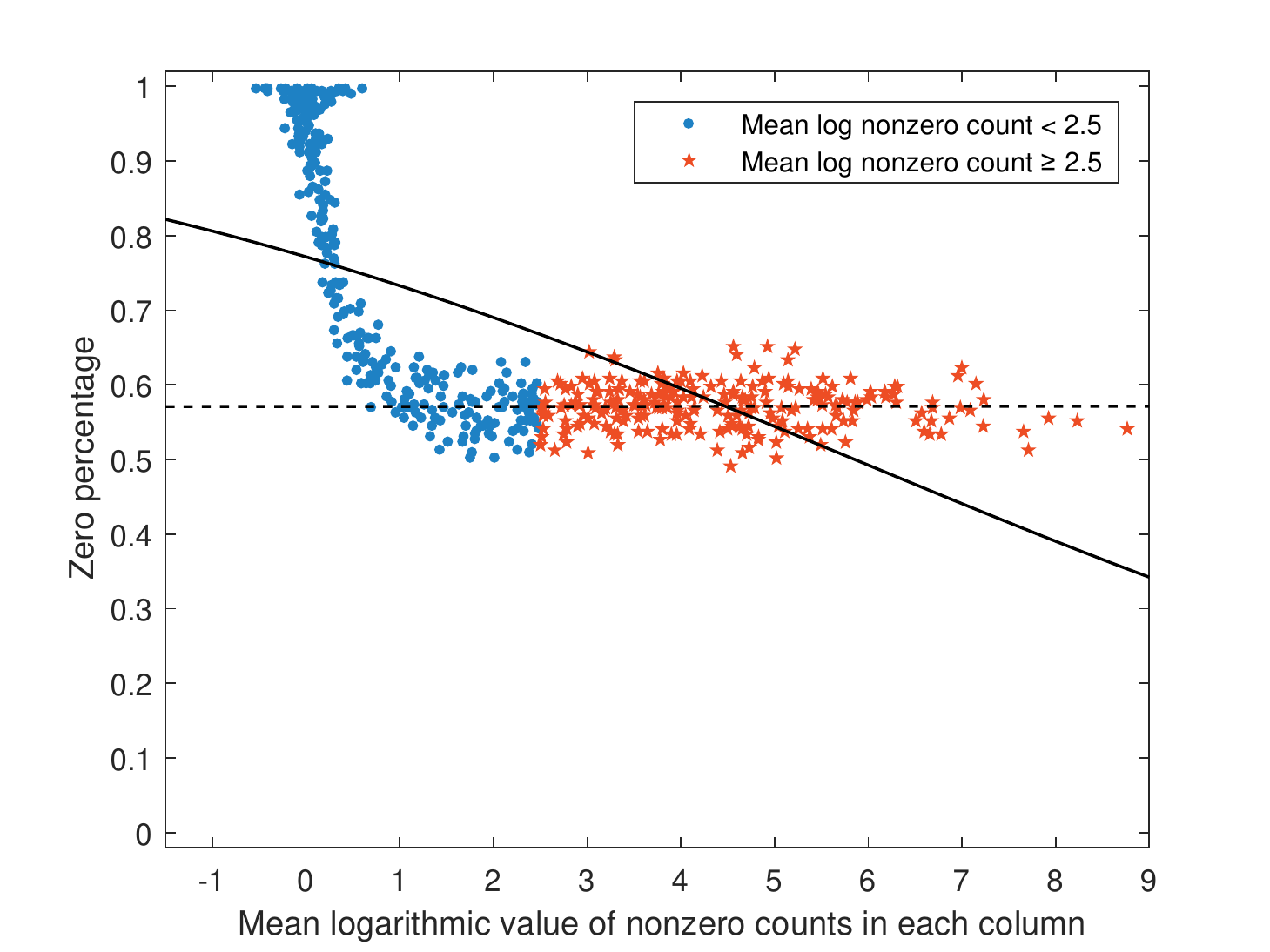}
		\caption{Relationship between zero probability and counts value of the simulation dataset. Zero percentage decreases first and then levels off at a fixed value. The red stars are the columns that have mean log nonzero counts large than 2.5 and the blue dots are other columns. The dashed curve is the fitted logistic function for red stars and the solid curve is the fitted logistic function for all points. }
		\label{p_proportion_web}
\end{figure}

\begin{figure}[!p]
	\parbox{0.5\textwidth}{
	\textbf{(a)}\\
	\includegraphics[width=0.48\textwidth]{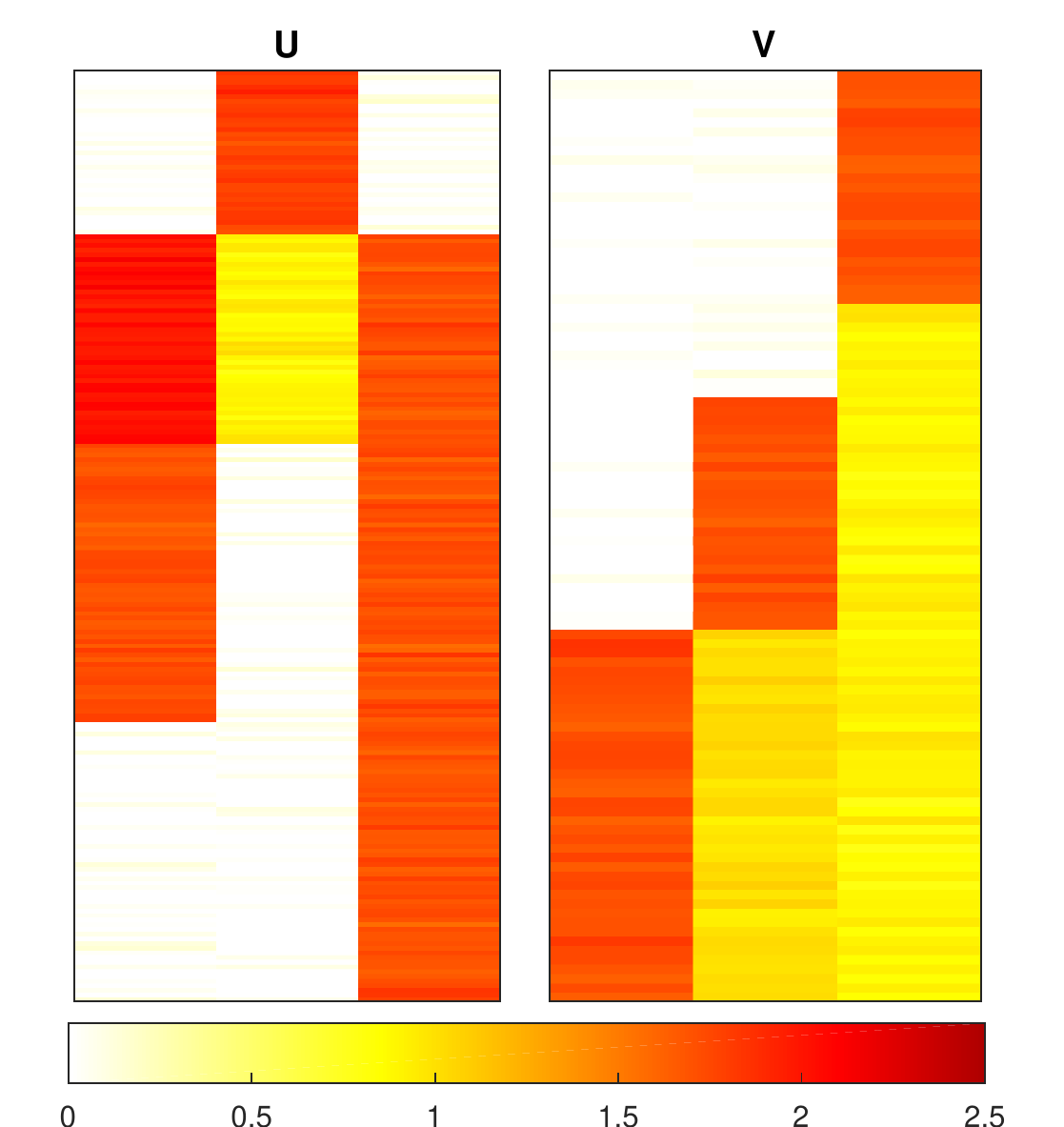}
	}
	\parbox{0.5\textwidth}{
	\textbf{(b)}\\
	\includegraphics[width=0.48\textwidth]{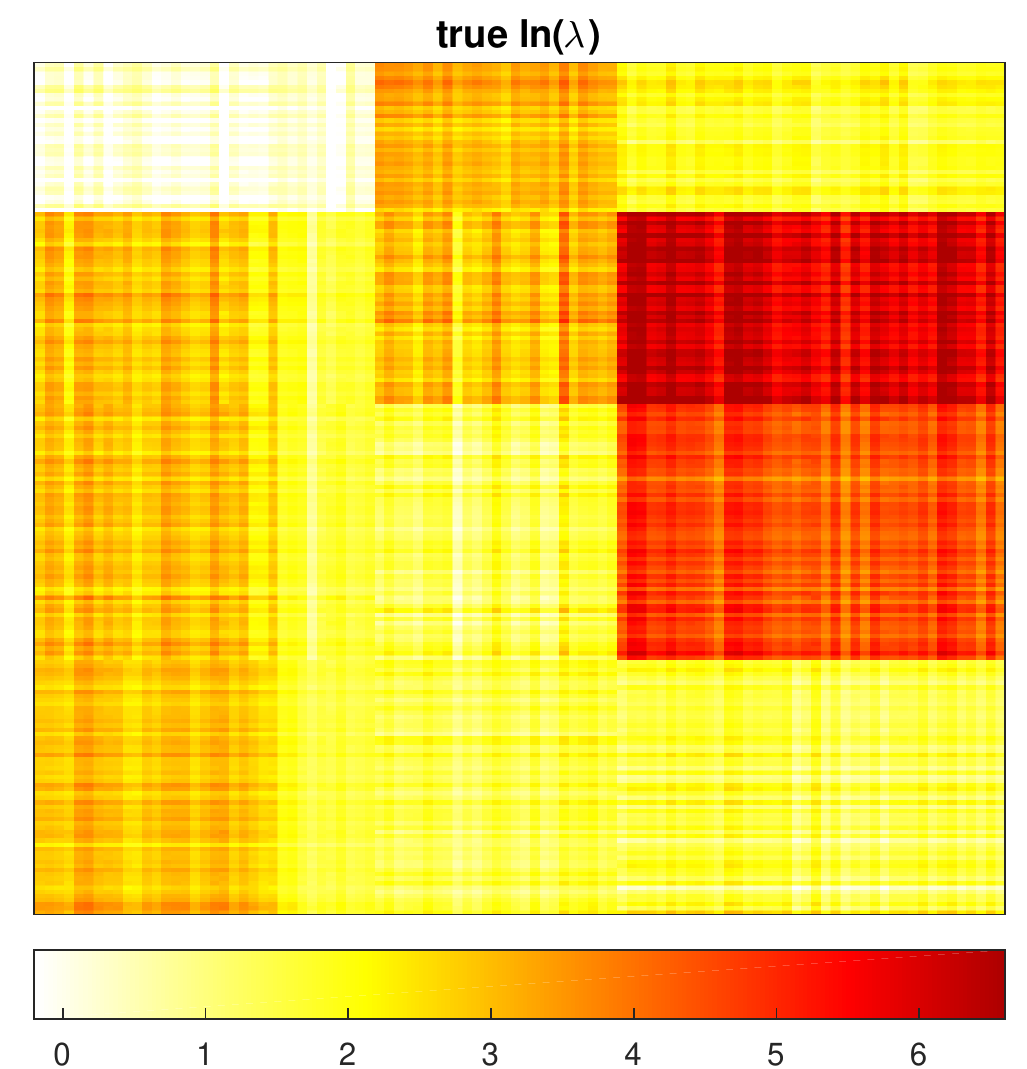}
	}
	\parbox{0.5\textwidth}{
	\textbf{(c)}\\
	\includegraphics[clip,trim=0cm 0.0cm 0.5cm 0cm,width=0.48\textwidth]{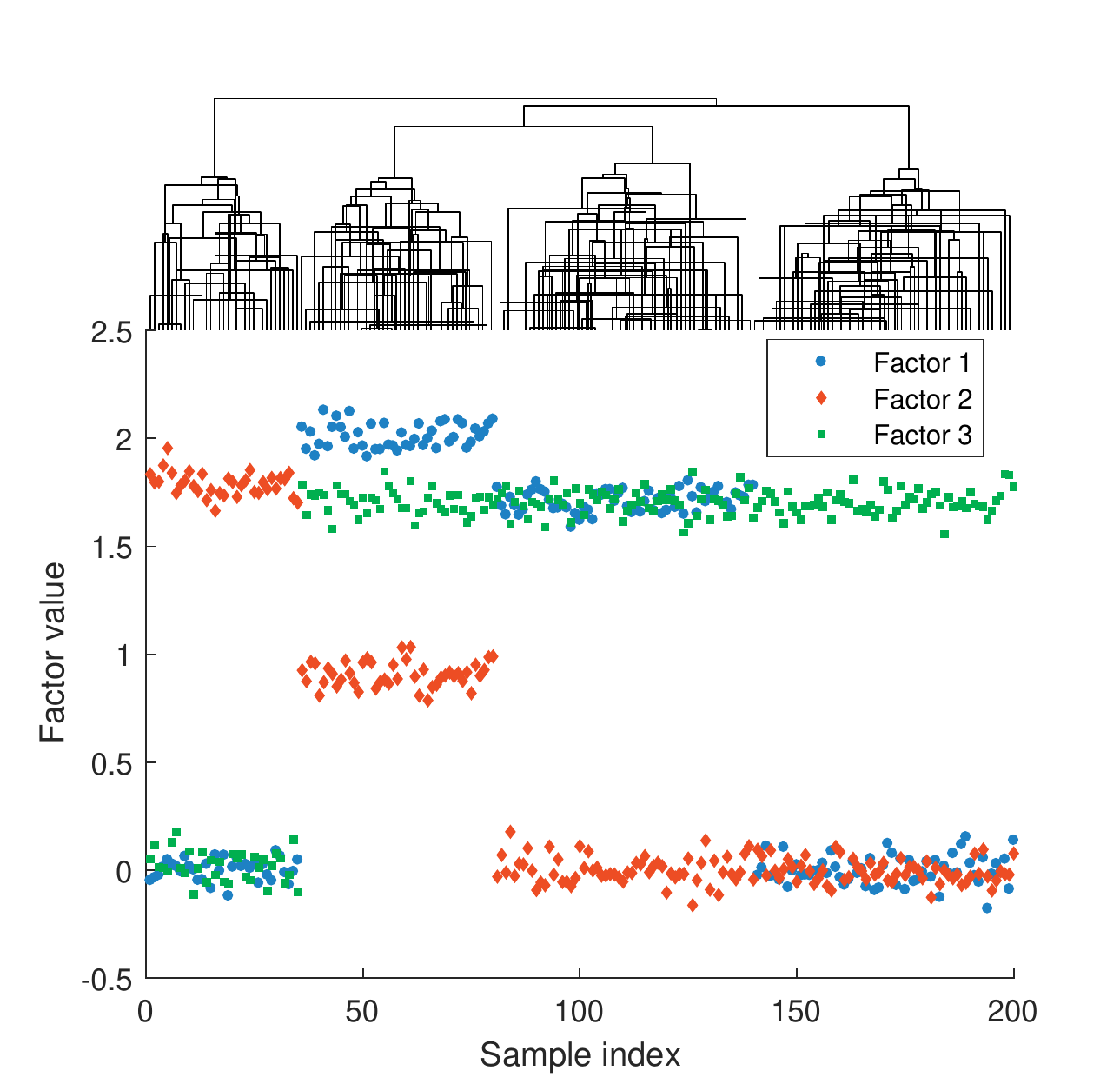}
	}
	\parbox{0.5\textwidth}{
	\textbf{(d)}\\
	\includegraphics[clip,trim=0cm 0.0cm 0.5cm 0cm,width=0.48\textwidth]{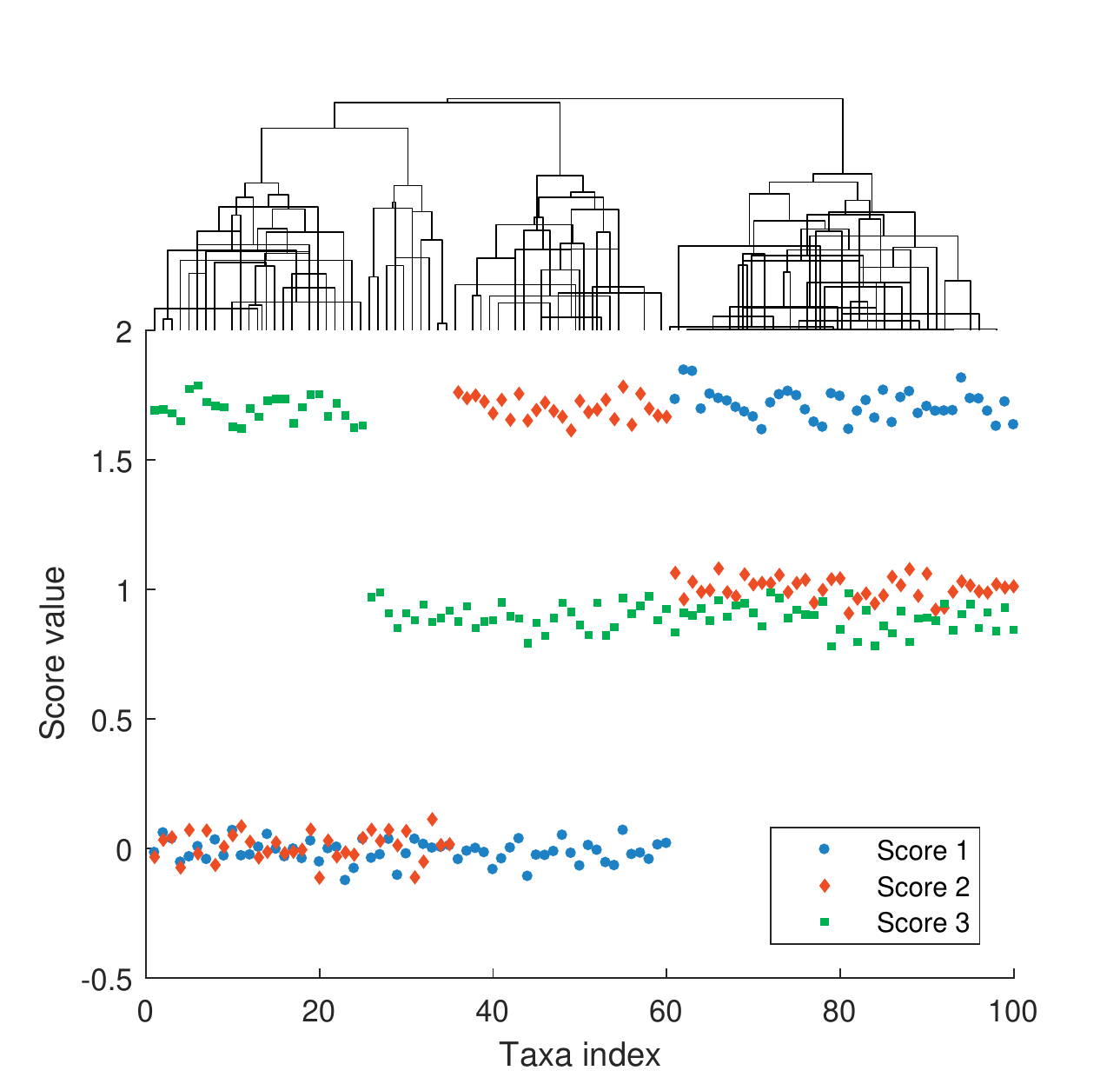}
	}
		\caption{Plots of simulation parameters. (a) True left singular vectors $U$ and true right singular vector $V$, indicating taxon clusters. (b) Heatmap of true $\ln(\lambda)$ matrix. (c)(d) The factor values for each sample/taxa. They could be clustered into 4 groups.}
		\label{u_v}
\end{figure}

{\begin{figure}[!ht]
	\parbox{1.05\textwidth}{\hspace{3.5ex} \textbf{ZIPFA} \hspace{16.5ex} \textbf{log-SVD} \hspace{13.5ex} \textbf{PSVDOS} \hspace{13ex} \textbf{GOMMS}}
		\parbox{1.05\textwidth}{
	\textbf{0\%:}\\
		\includegraphics[clip,trim=0.8cm 0cm 0cm .5cm,width=0.25\textwidth]{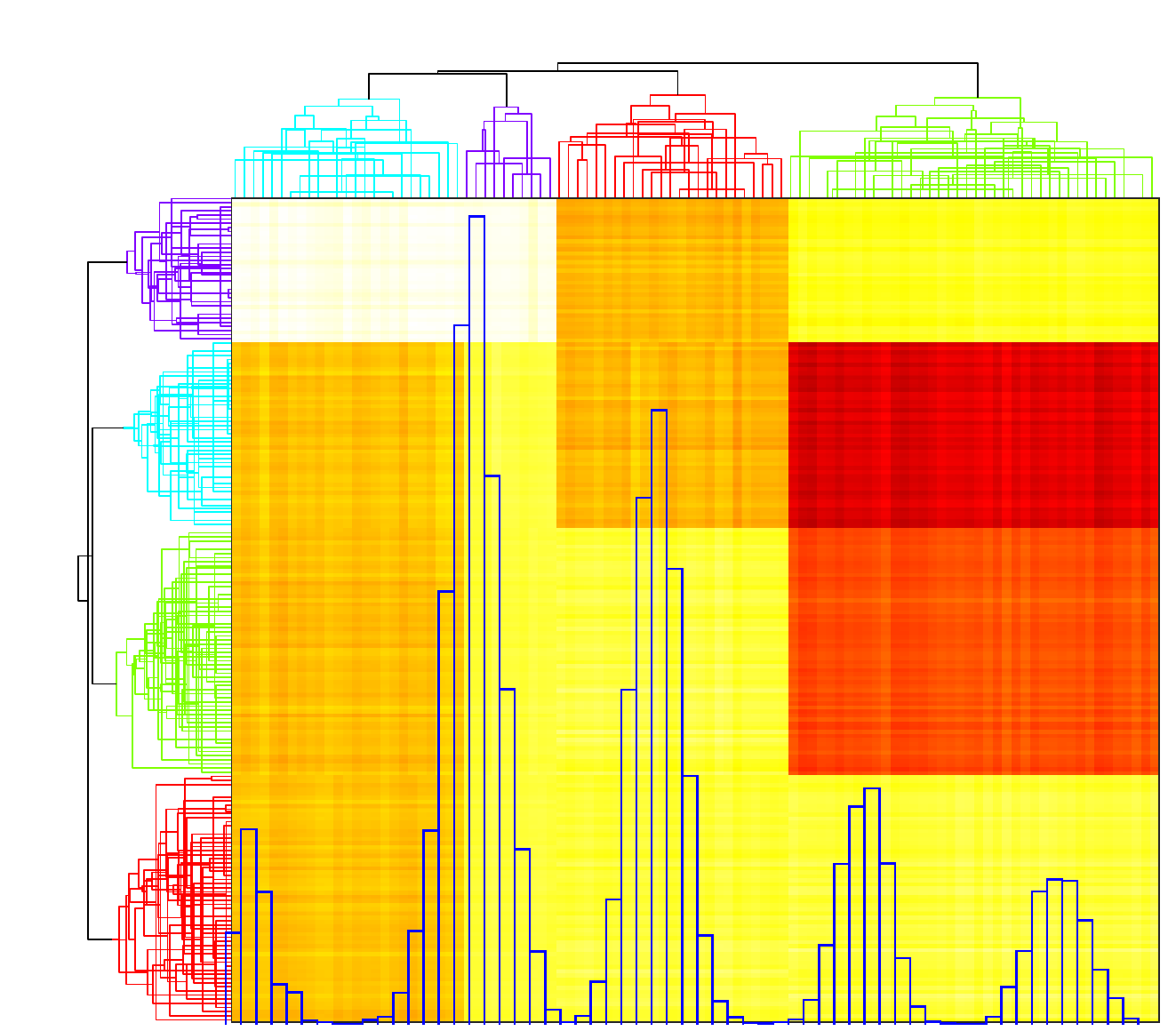}
	\includegraphics[clip,trim=0.8cm 0cm 0cm .5cm,width=0.25\textwidth]{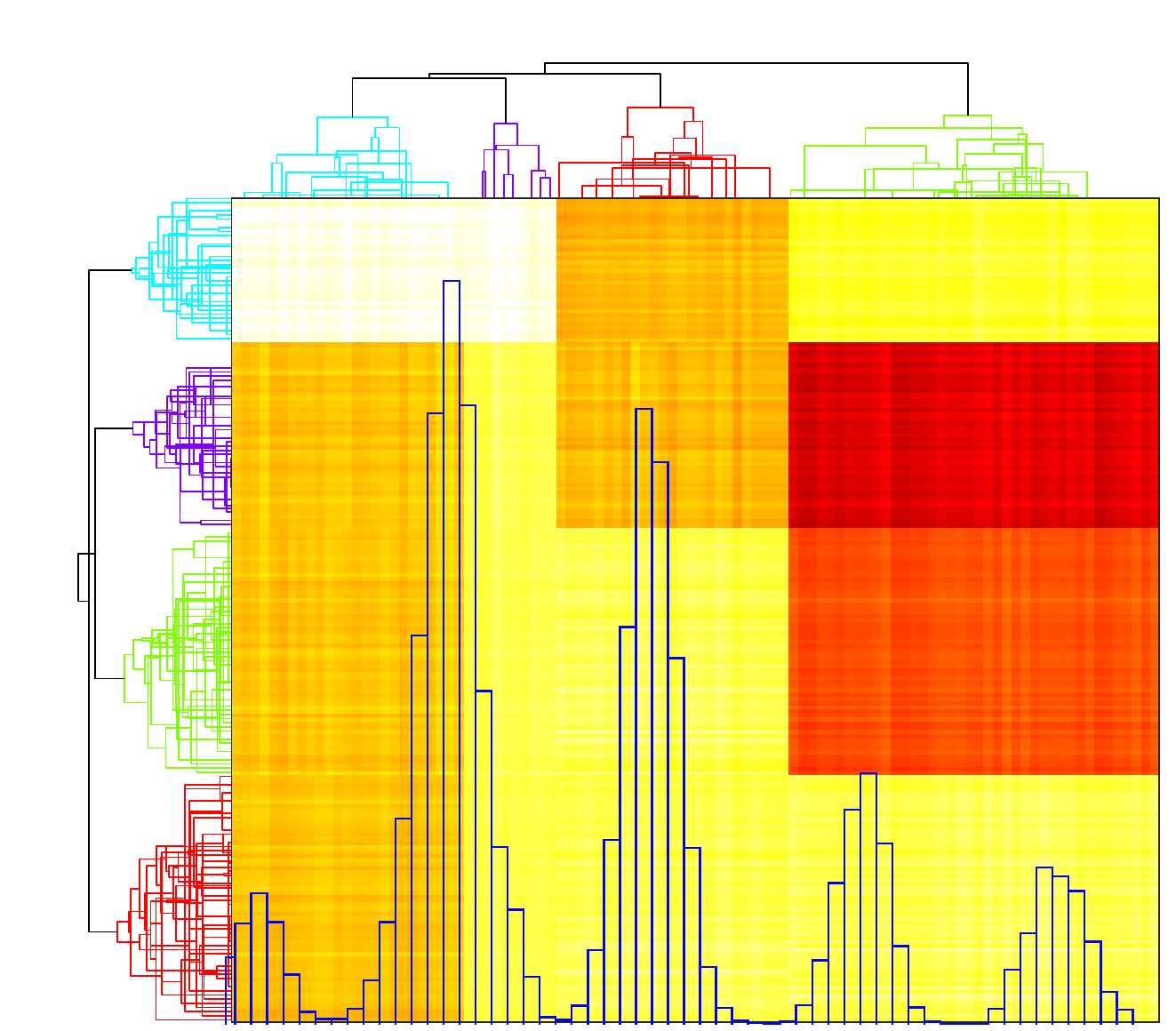}
	\includegraphics[clip,trim=0.8cm 0cm 0cm .5cm,width=0.25\textwidth]{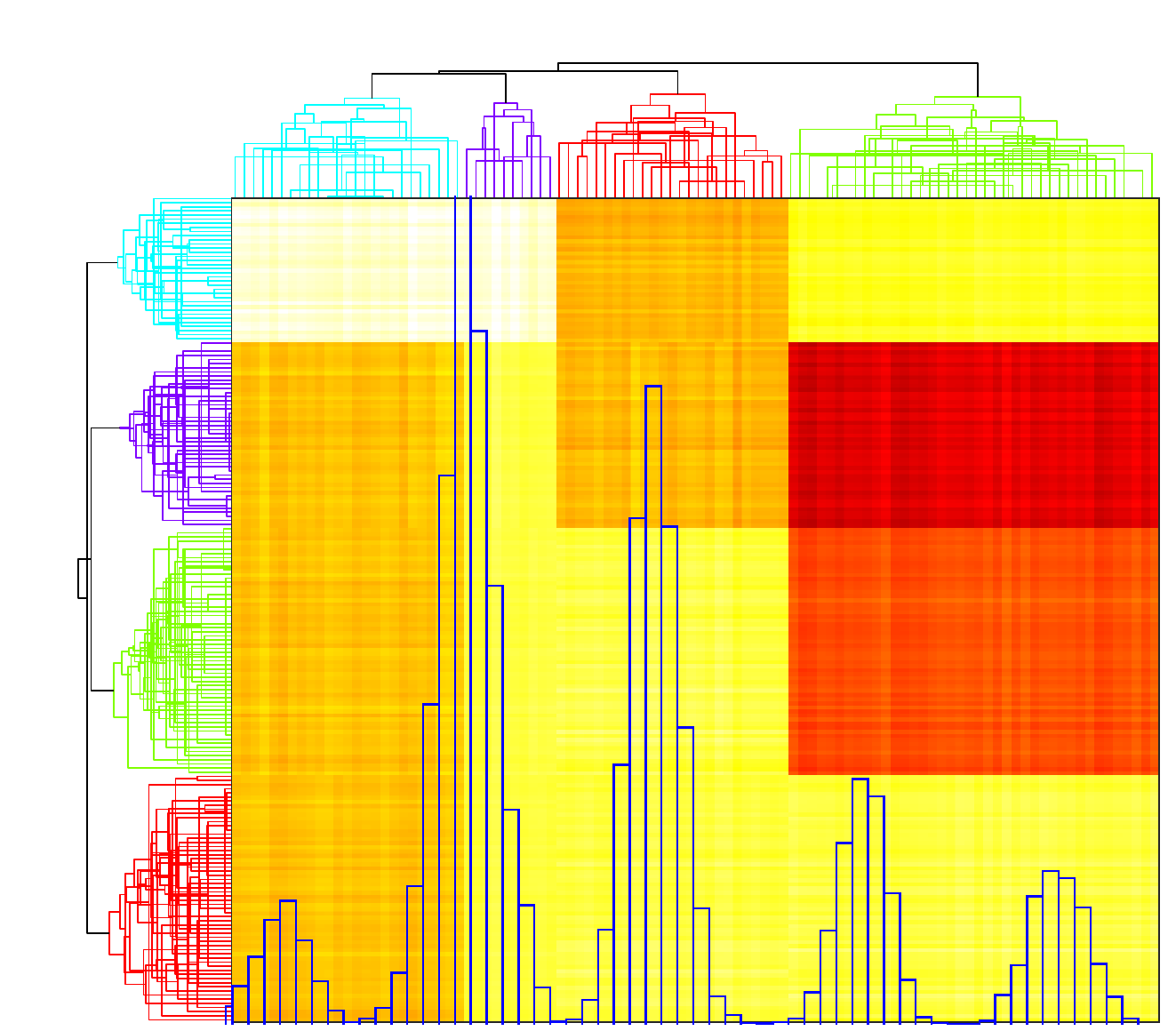}
	\includegraphics[clip,trim=0.8cm 0cm 0cm .5cm,width=0.25\textwidth]{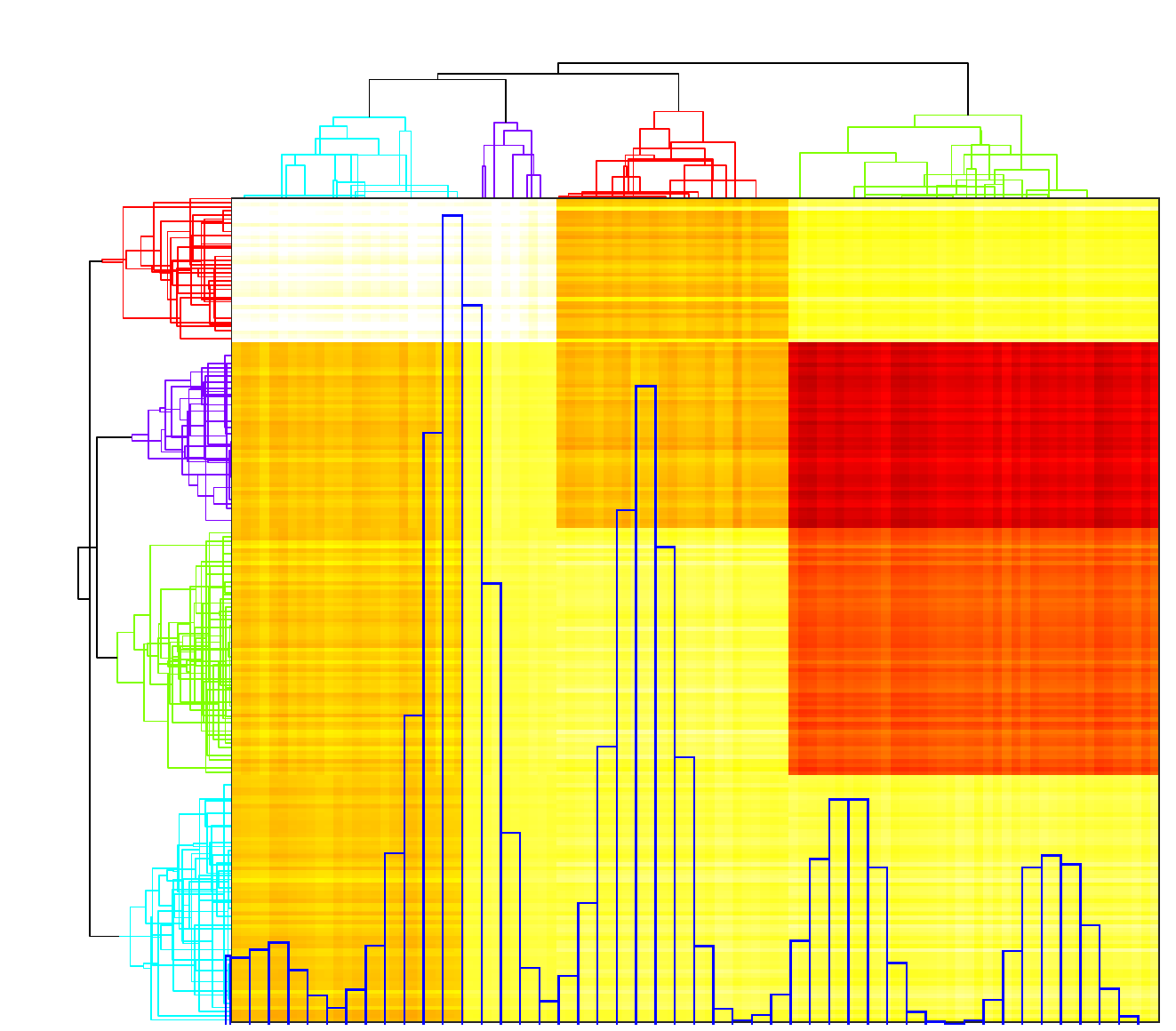}
		}
		\parbox{1.05\textwidth}{
		\textbf{20\%:}\\
		\includegraphics[clip,trim=0.8cm 0cm 0cm .5cm,width=0.25\textwidth]{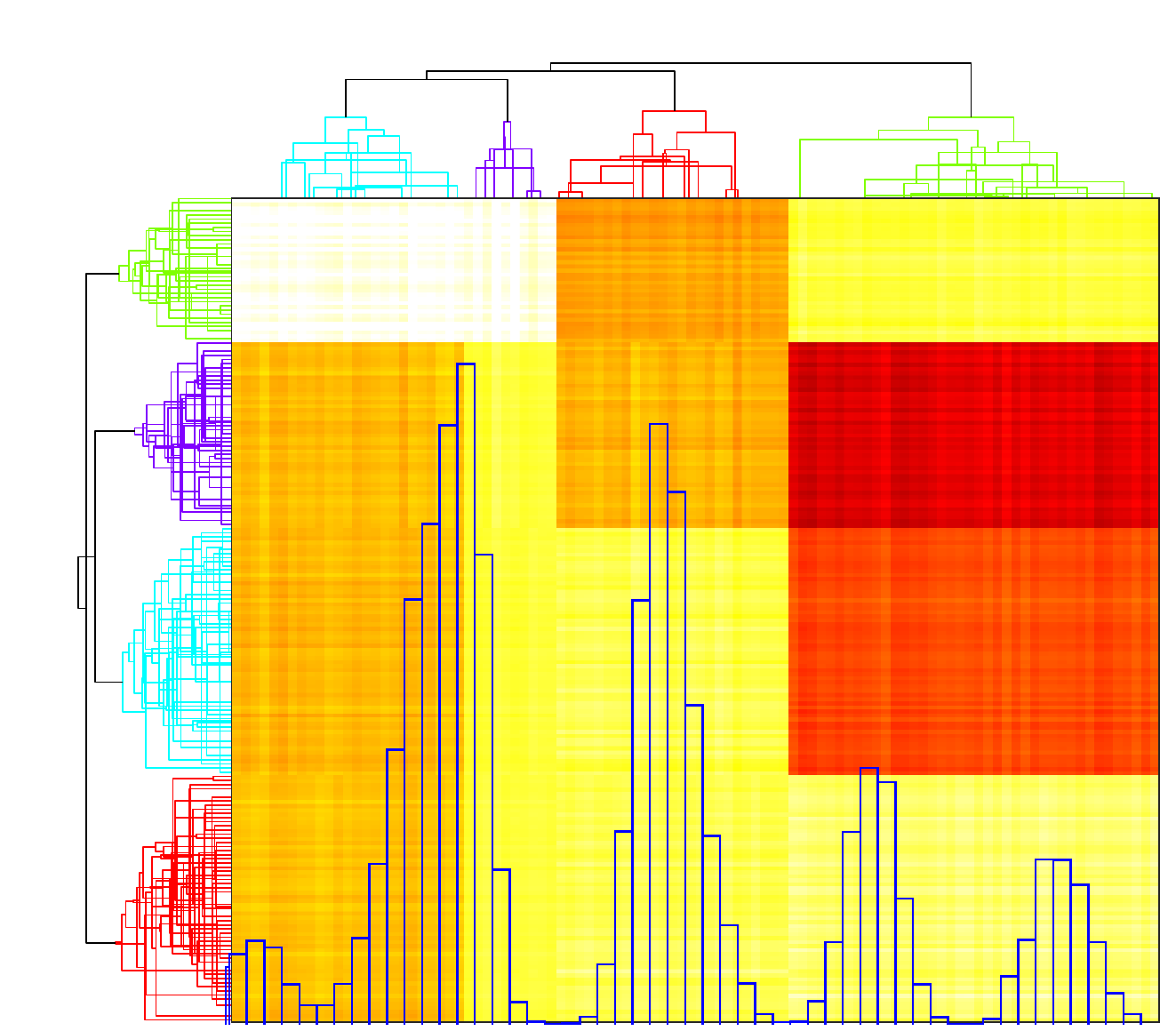}
	\includegraphics[clip,trim=0.8cm 0cm 0cm .5cm,width=0.25\textwidth]{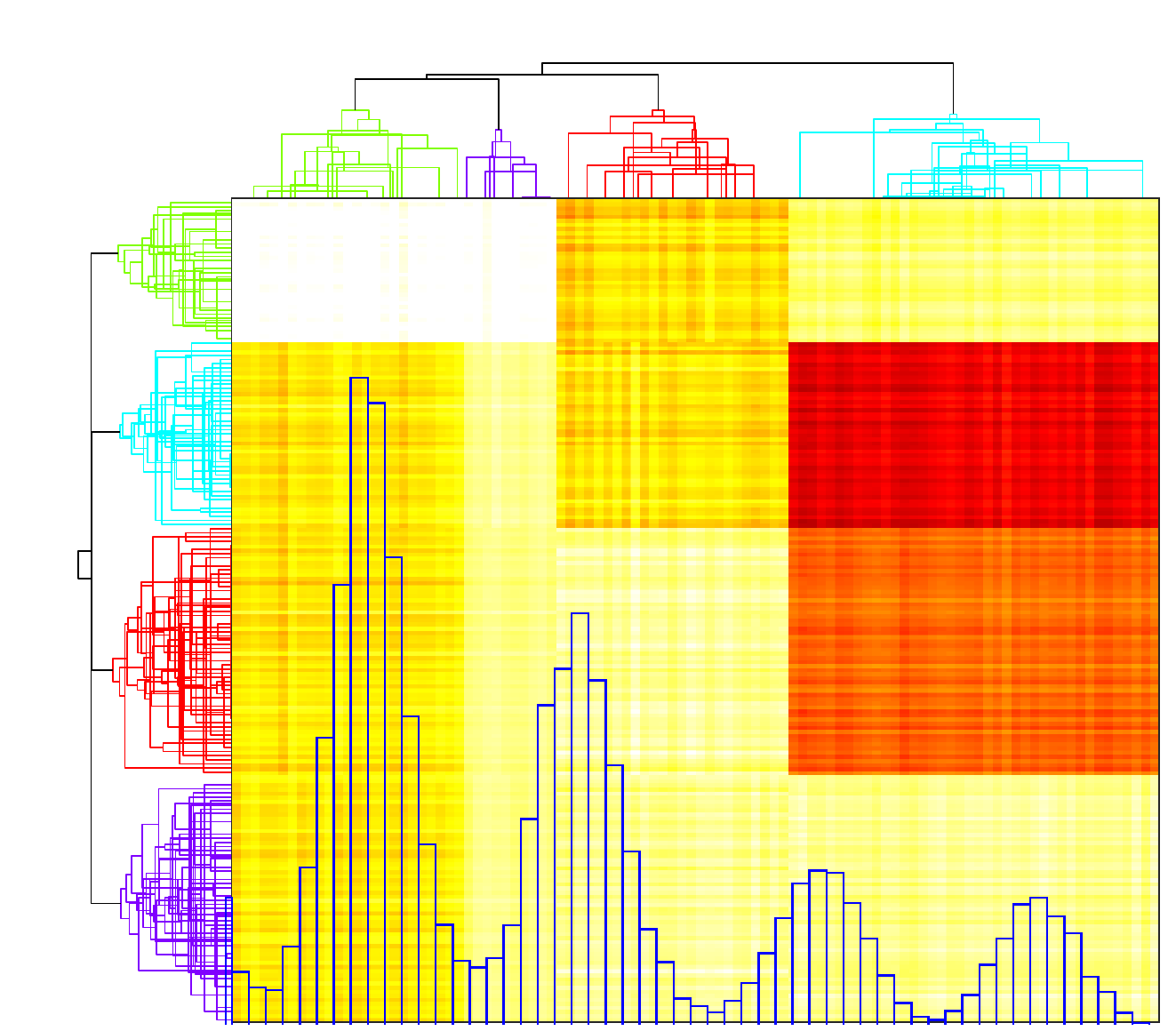}
	\includegraphics[clip,trim=0.8cm 0cm 0cm .5cm,width=0.25\textwidth]{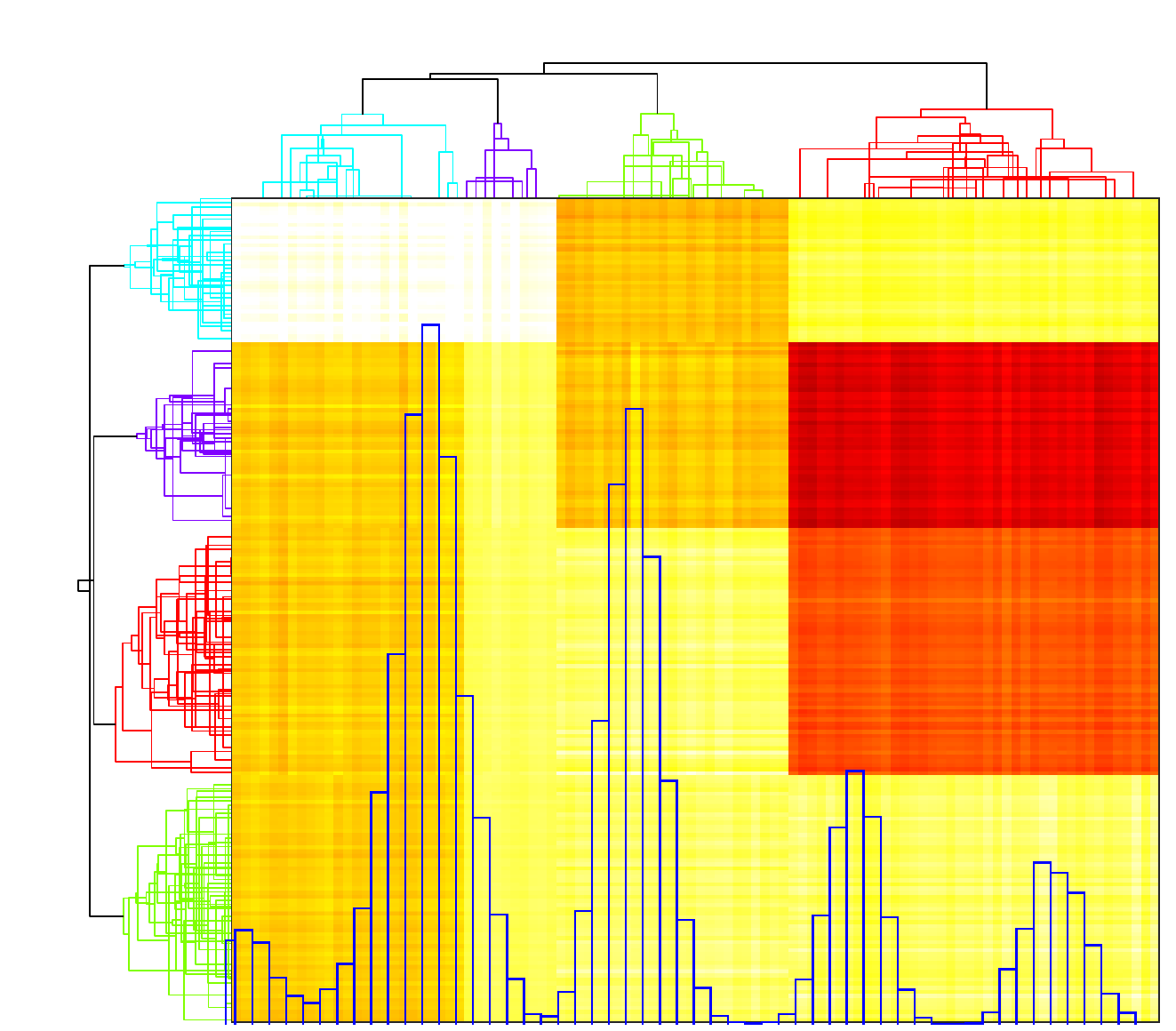}
	\includegraphics[clip,trim=0.8cm 0cm 0cm .5cm,width=0.25\textwidth]{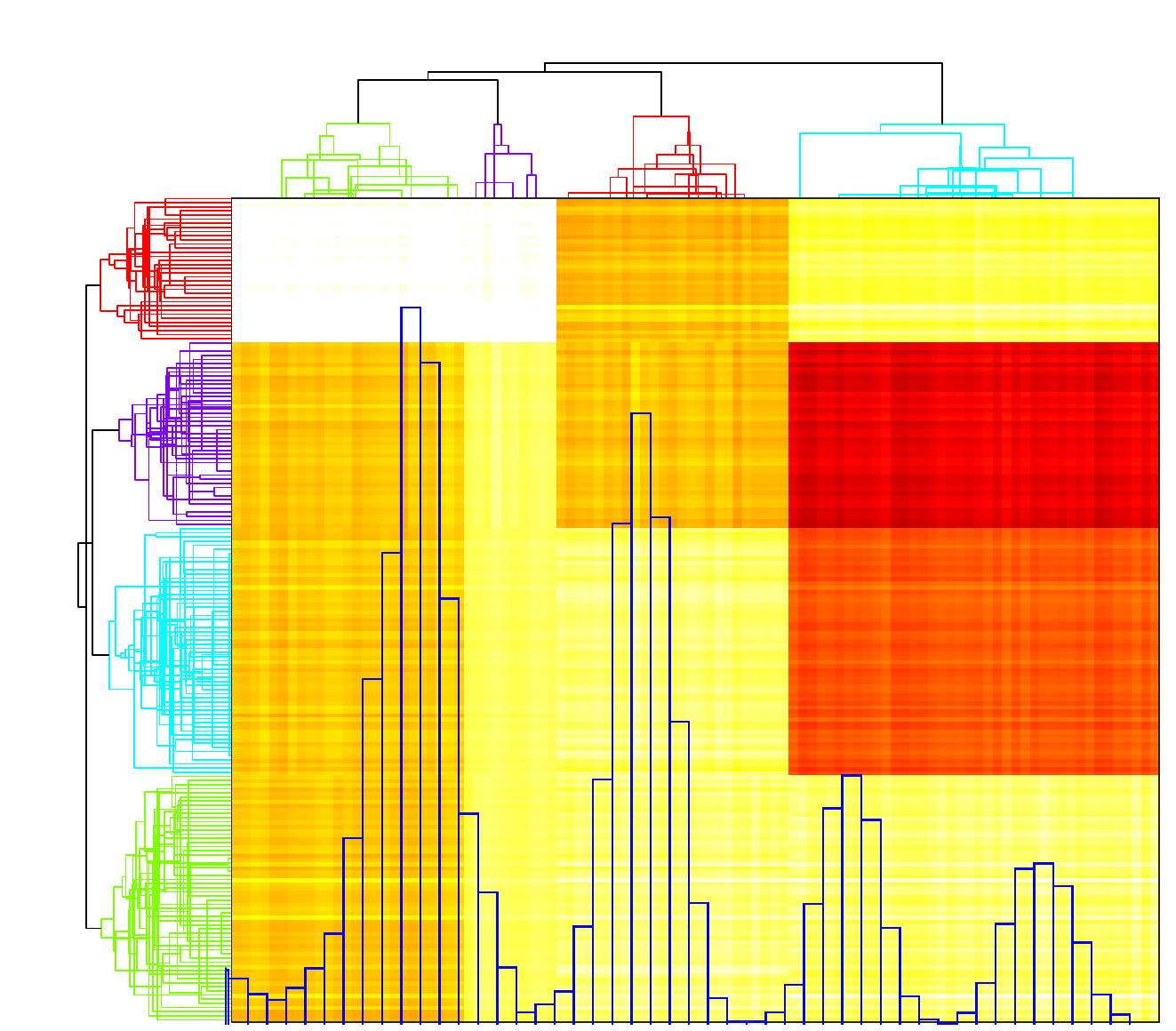}
		}
			\parbox{1.05\textwidth}{
			\textbf{40\%:}\\
		\includegraphics[clip,trim=0.8cm 0cm 0cm .5cm,width=0.25\textwidth]{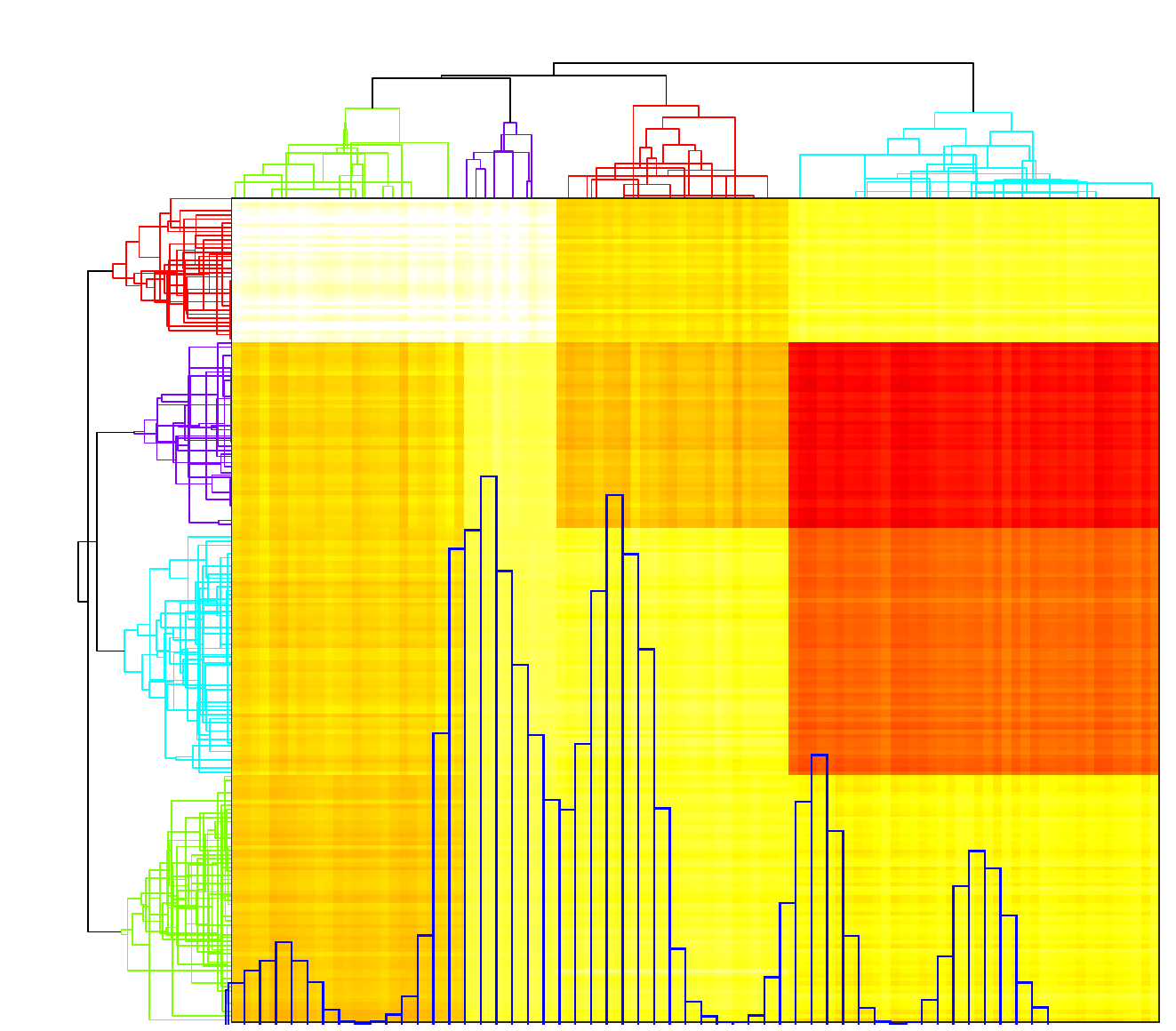}
	\includegraphics[clip,trim=0.8cm 0cm 0cm .5cm,width=0.25\textwidth]{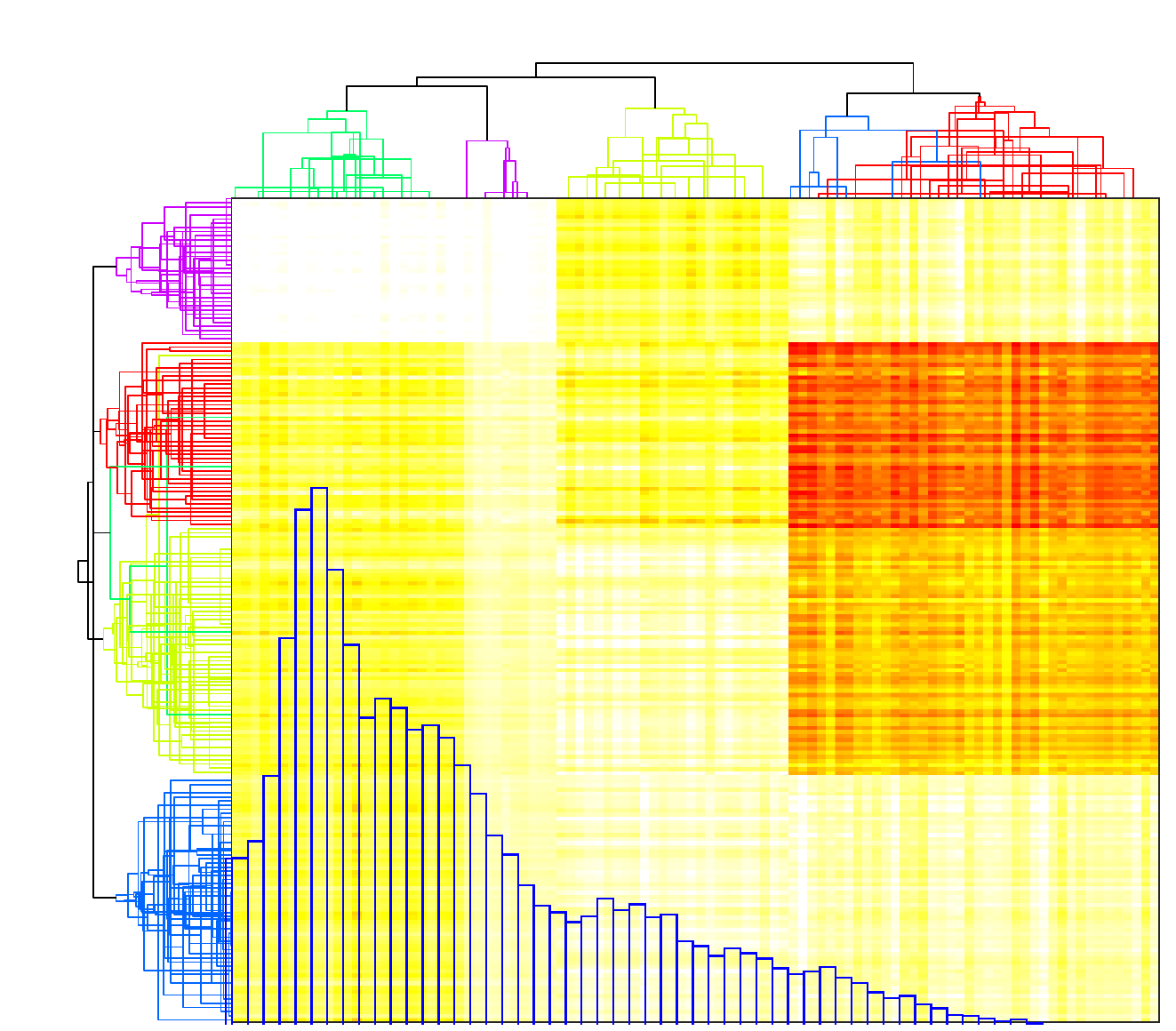}
	\includegraphics[clip,trim=0.8cm 0cm 0cm .5cm,width=0.25\textwidth]{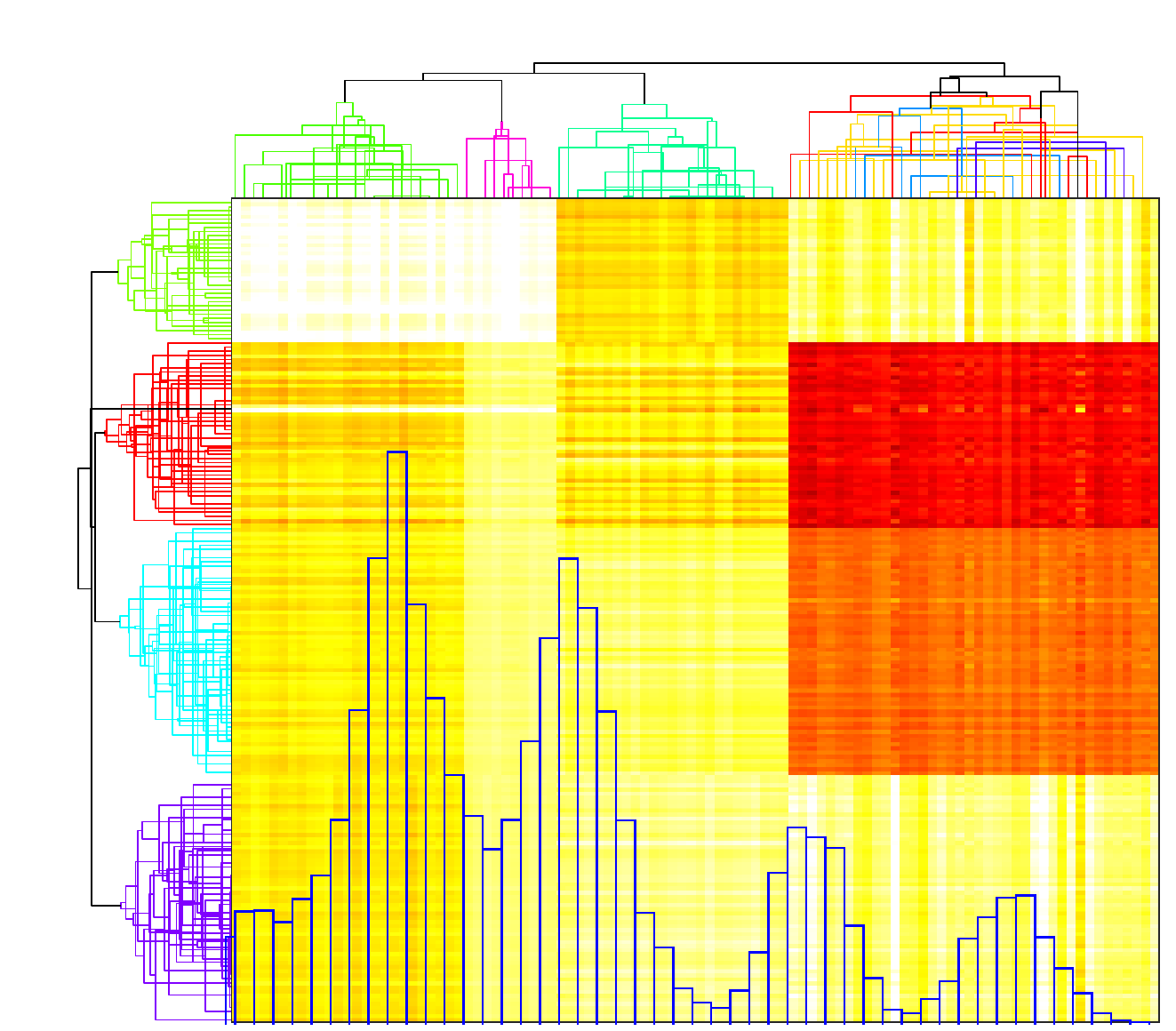}
	\includegraphics[clip,trim=0.8cm 0cm 0cm .5cm,width=0.25\textwidth]{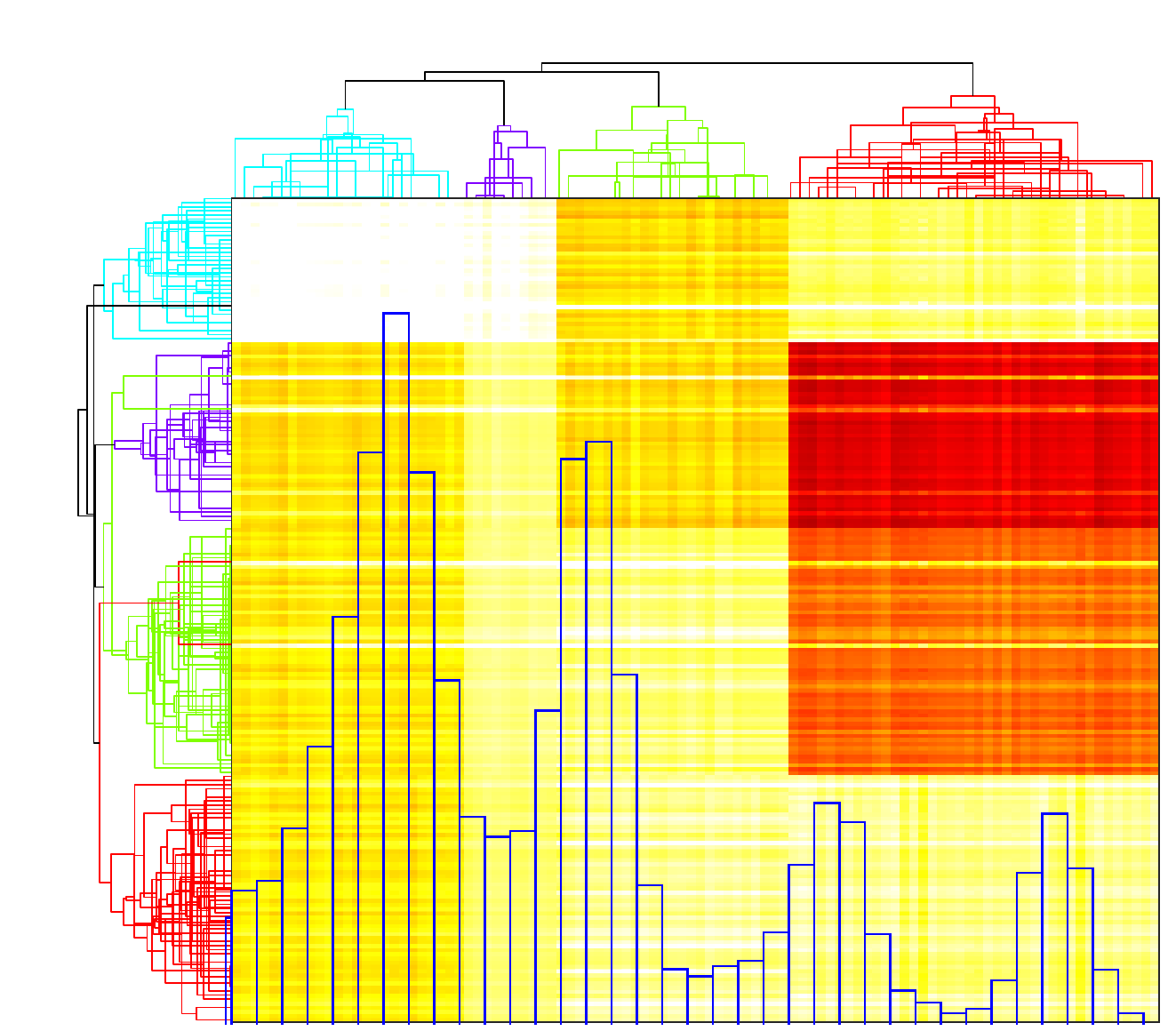}
		}
		\caption{Heatmap of $\hat U\hat V^\top$ from different methods in setting 1. Blue histogram shows the distribution of fitted $\hat U\hat V^\top$; Phylogenetic tree on the top and left shows clustering of taxa and samples.}
		\label{fig:different_method}
\end{figure}

\begin{figure}[!ht]
	\centering	\includegraphics[clip,trim= 0 0 1cm .3cm,width=0.6\textwidth]{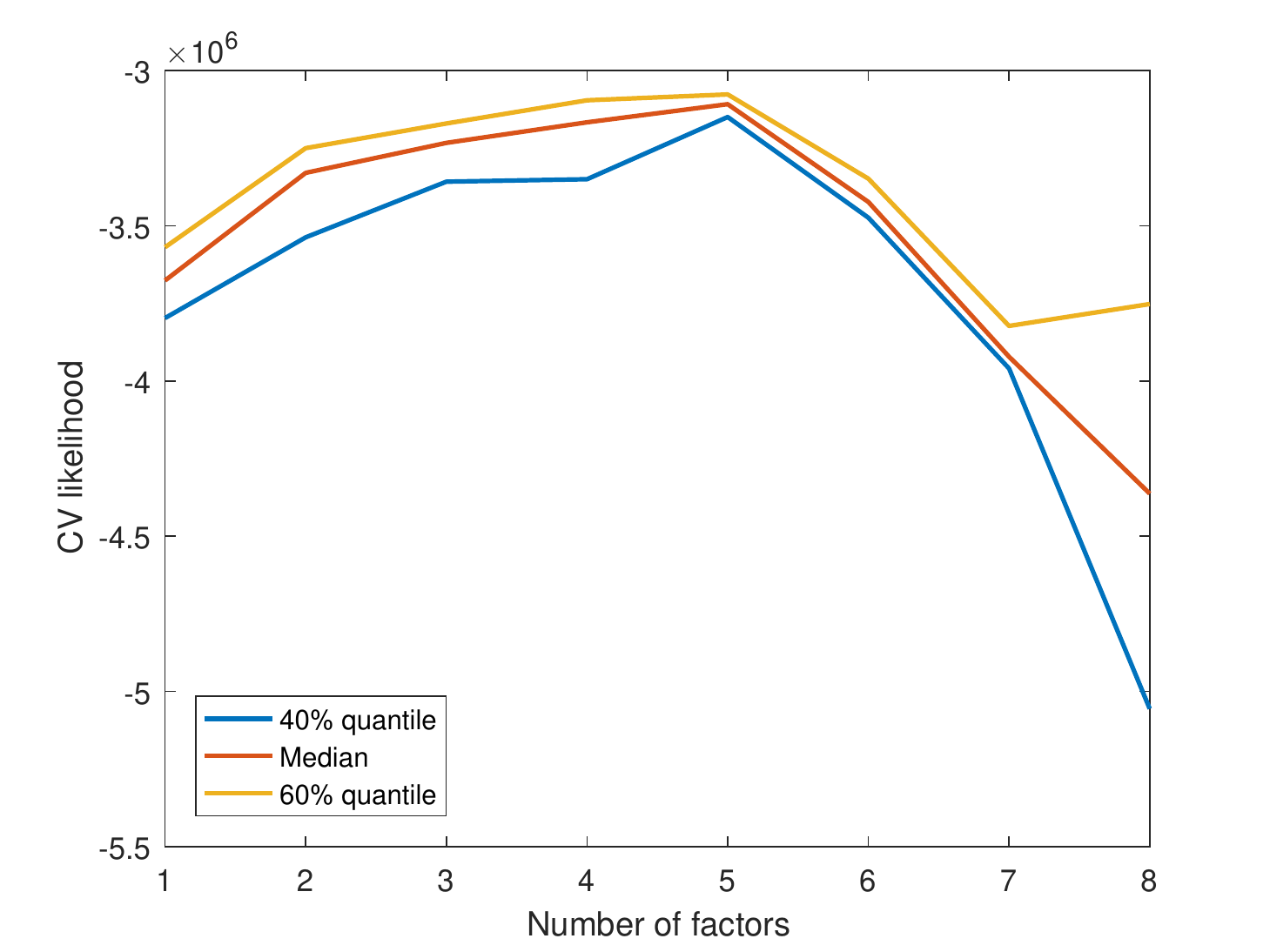}
	\caption{Cross validation to choose the number of factors in real data analysis. At a specified rank, we obtain several CV likelihoods by assigning test sets randomly. Three solid lines indicate 40\%, 50\% and 60\% quantile of the CV likelihoods. }
	\label{fig:cv}
\end{figure}

\begin{figure}[!ht]
	\centering	\includegraphics[clip,trim= 0 0 1cm .3cm,width=0.6\textwidth]{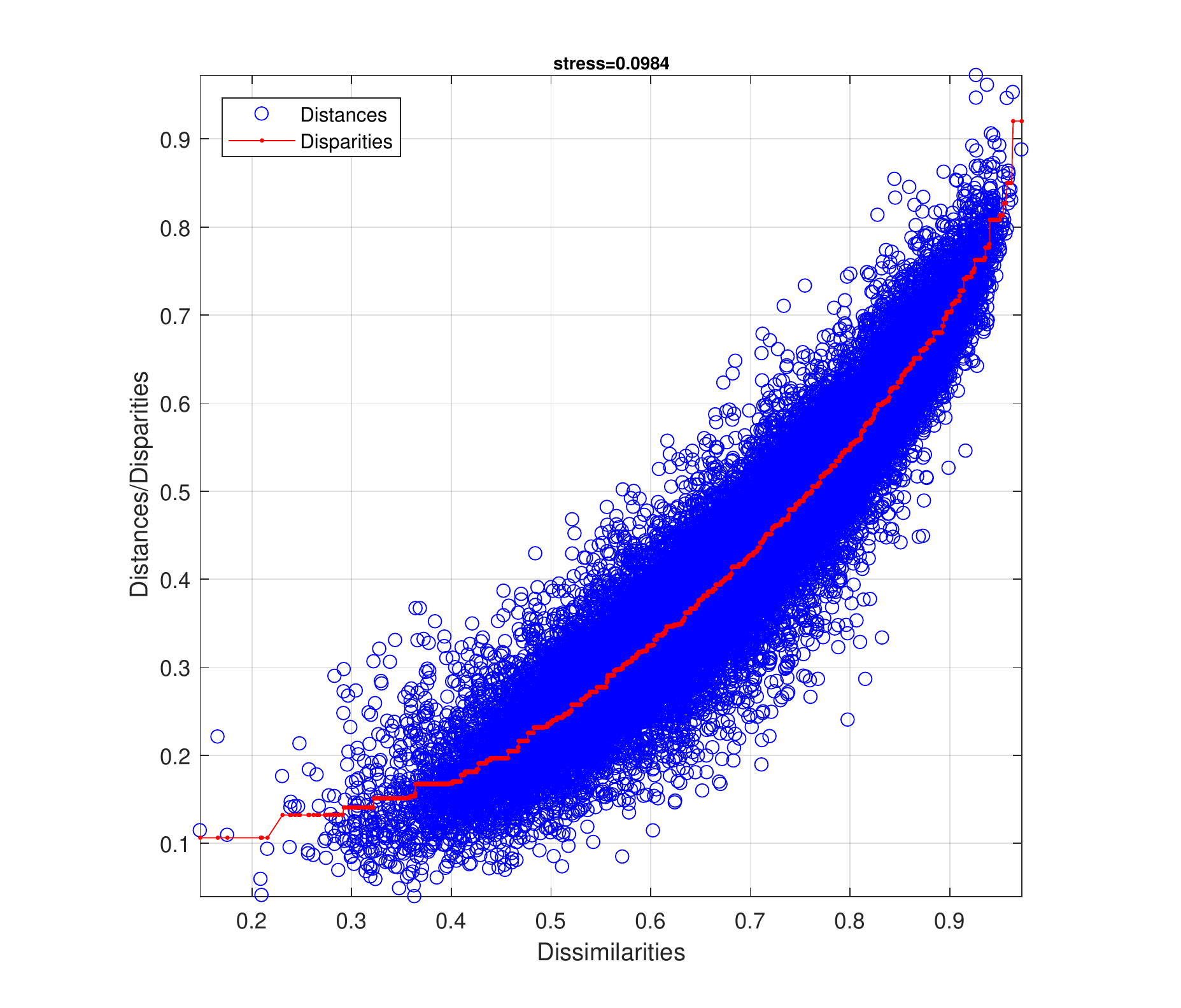}
	\caption{Shepard stress plot for nMDS model in Subsection 4.2}
	\label{fig:stress}
\end{figure}

\begin{figure}[!p]
	\parbox{0.5\textwidth}{
	\textbf{(a)}\\
	\includegraphics[clip,trim=.6cm 0.0cm 0.7cm .6cm,width=0.5\textwidth]{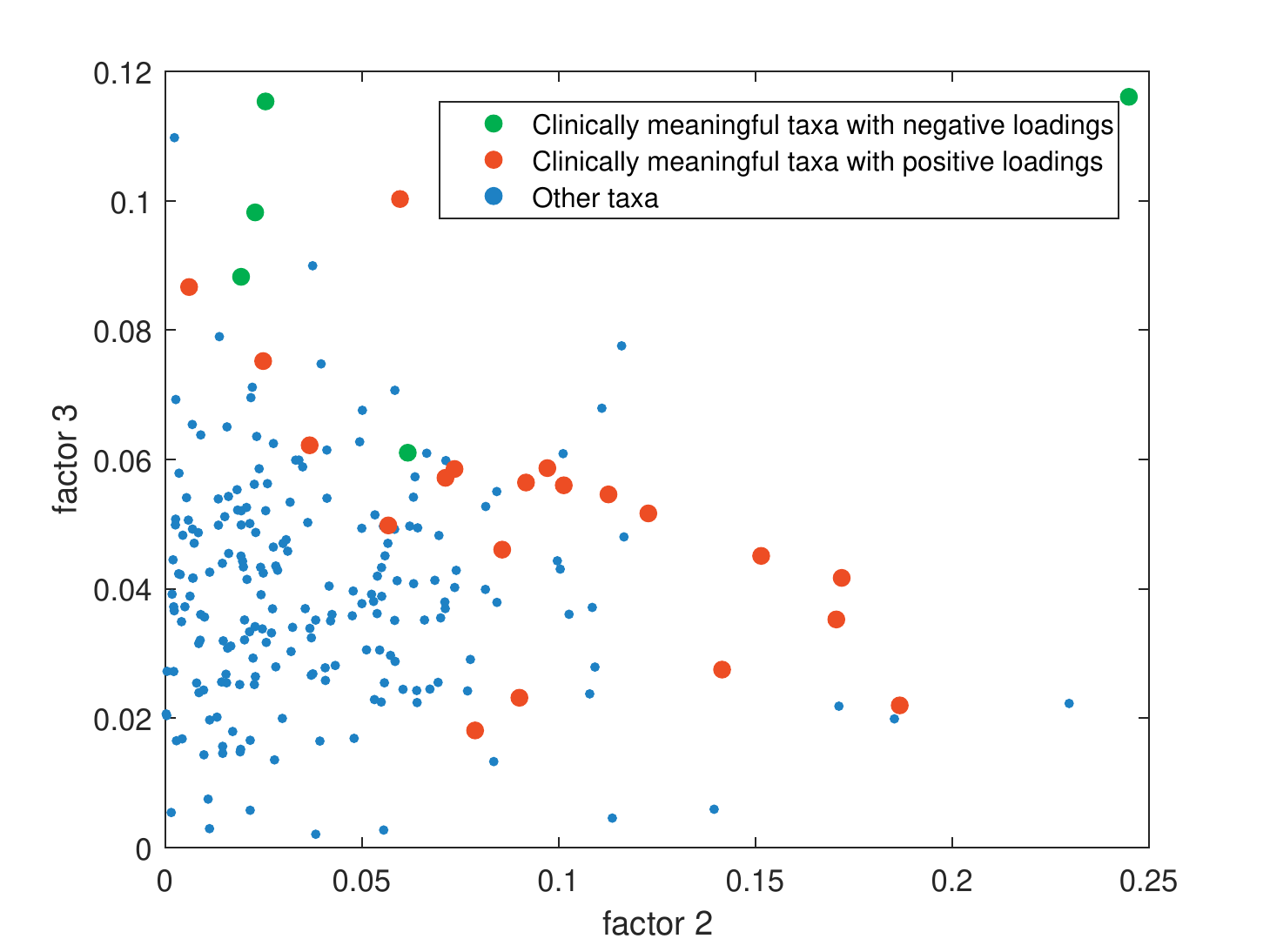}
	}
	\parbox{0.5\textwidth}{
	\textbf{(b)}\\
	\includegraphics[clip,trim=.6cm 0.0cm 0.7cm .6cm,width=0.5\textwidth]{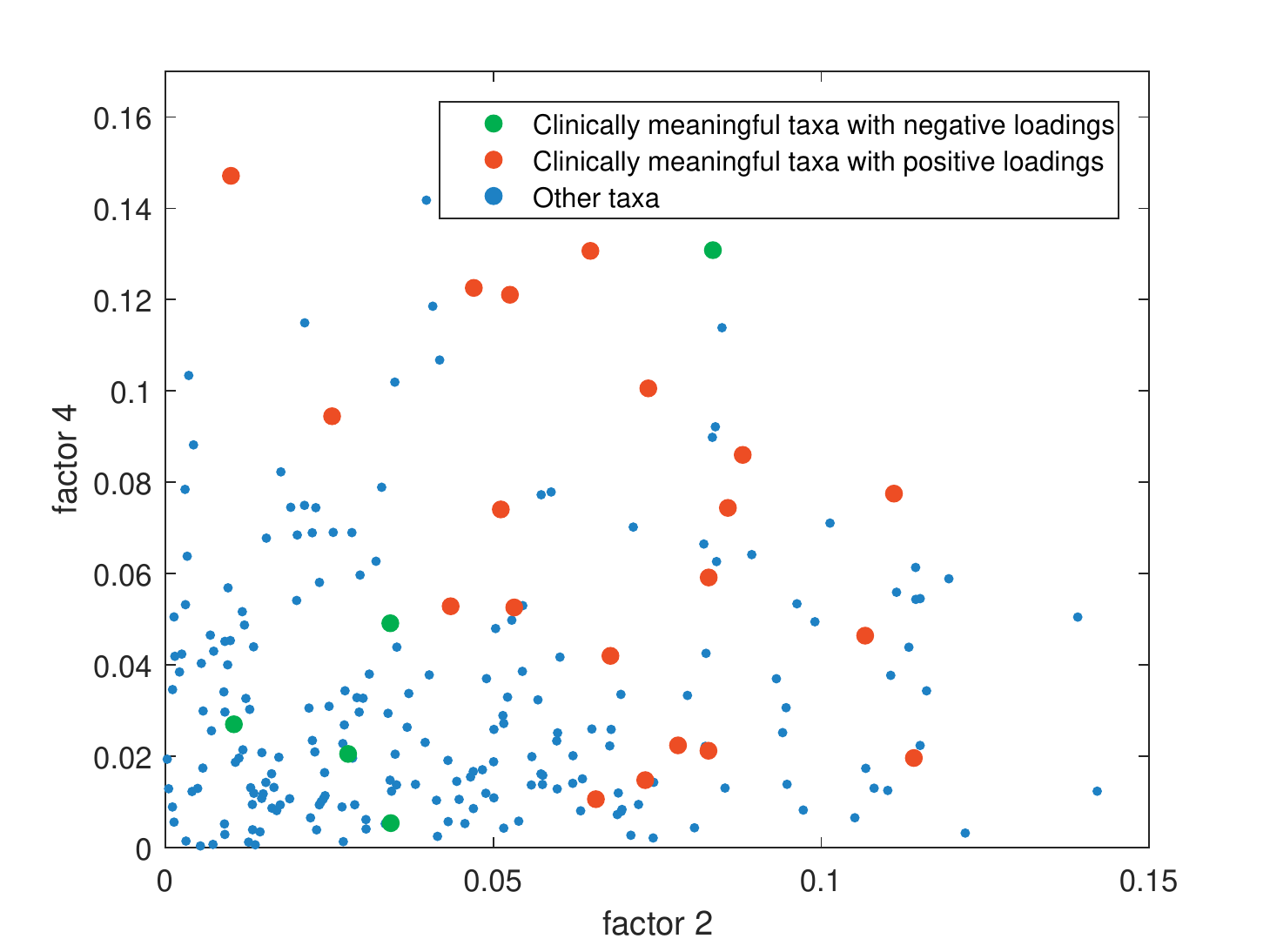}
	}
	\parbox{0.5\textwidth}{
	\textbf{(c)}\\
	\includegraphics[clip,trim=.6cm 0.0cm 0.7cm .6cm,width=0.5\textwidth]{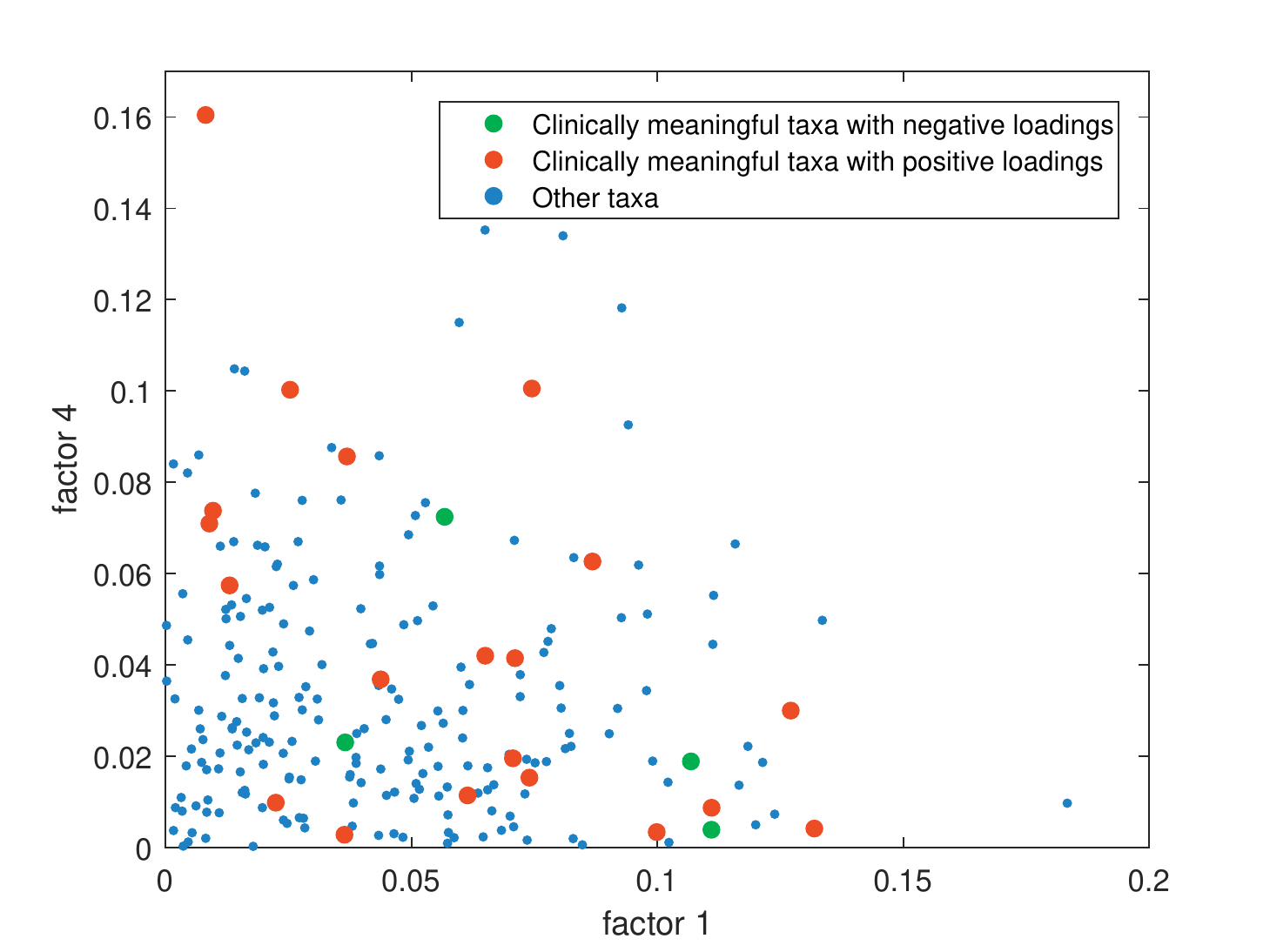}
	}
	\parbox{0.5\textwidth}{
	\textbf{(d)}\\
	\includegraphics[clip,trim=.6cm 0.0cm 0.7cm .6cm,width=0.5\textwidth]{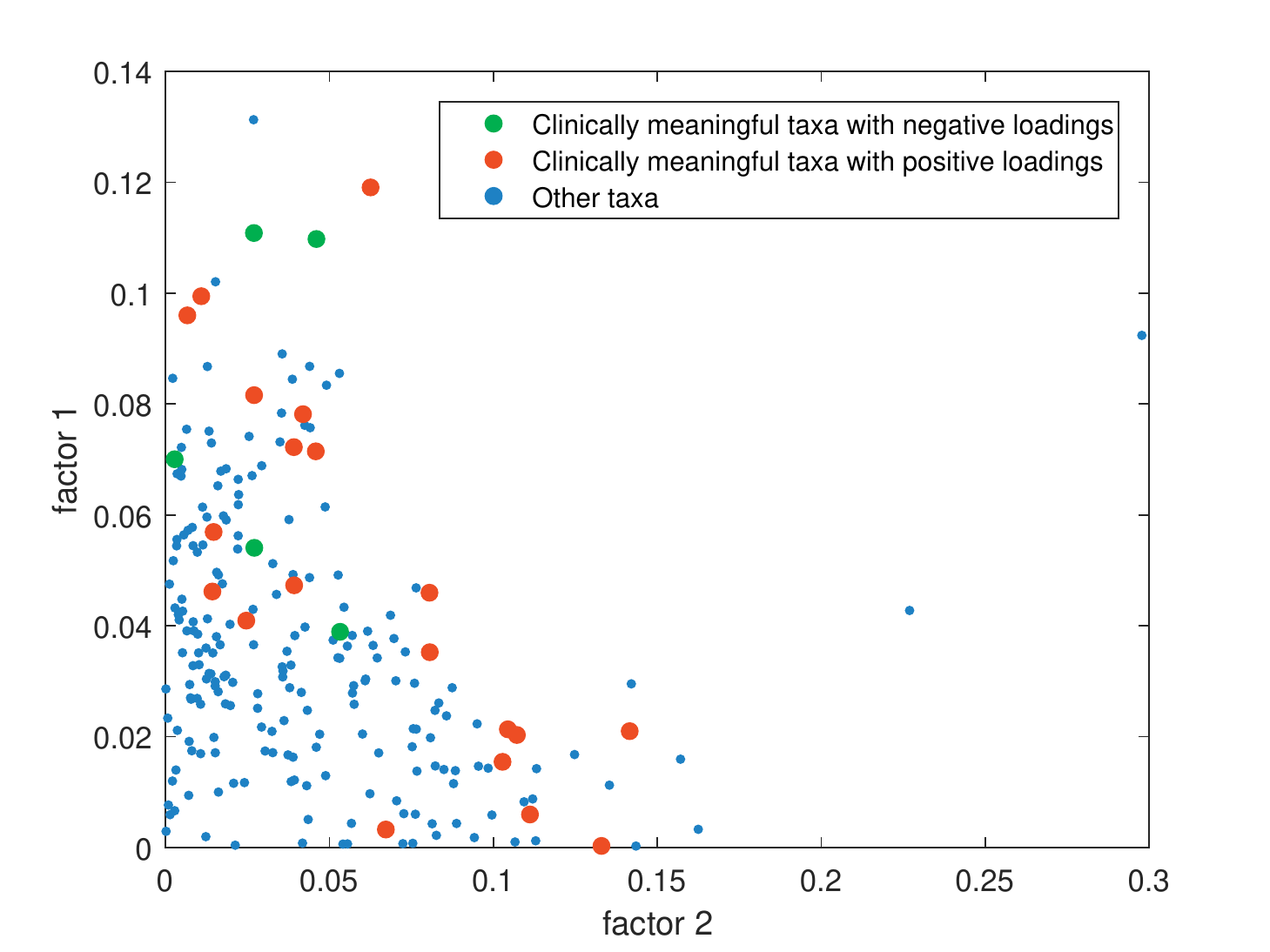}
	}
		\caption{Absolute taxa loadings on the two most significant factors. Each point is a loading coordinate of a taxon on these two factors. Blue or red dots are clinically meaningful taxa in other literature. (a) Loadings on factor 2, 3 of our proposed ZIPFA. (b) Loadings on factor 2, 4 of log-SVD. (c) Loadings on factor 1, 4 of PSVDOS. (d) Loadings on factor 1, 2 of GOMMS}
\end{figure}

\begin{sidewaystable}[!ht]
\begin{minipage}{1.090\textwidth}
		\caption{Coefficients and p-value of 5 factors in different models}
		\label{pvalue}
		\centering
		\newcolumntype{Y}{>{\centering\arraybackslash}X}
		\hspace{0pt}
\begin{tabularx}{\textwidth}{cYYYYYYYYYYYY}
	\toprule
	\multirow{2}[2]{*}{Response} & \multicolumn{2}{c}{ZIPFA} & \multicolumn{2}{c}{log-SVD} & \multicolumn{2}{c}{PSVDOS} & \multicolumn{2}{c}{GOMMS} & \multicolumn{2}{c}{PCoA} & \multicolumn{2}{c}{nMDS} \\
			 & coef     & p-value  & coef     & p-value  & coef     & p-value  & coef     & p-value  & coef     & p-value  & coef     & p-value \\
	\midrule
	\multirow{5}[0]{*}{MeanAttLoss \footnote{Mean attachment loss.}} & 0.0215   & 0.038\dottt   & 0.0039   & 0.434    & 0.0043   & 0.352    & 0.0009   & 0.903    & -0.4522  & 0.008\dotttt  & -0.5448  & 0.004\dotttt \\
			 & -0.0039  & 0.029\dottt   & 0.0090   & 0.001\dotttt  & 0.0018   & 0.673    & -0.0034  & 0.150    & 0.1073   & 0.587    & -0.3279  & 0.154 \\
			 & 0.0138   & 0.003\dotttt  & -0.0053  & 0.190    & -0.0020  & 0.294    & 0.0020   & 0.492    & -0.4579  & 0.033\dottt   & -0.2522  & 0.295 \\
			 & 0.0071   & 0.089\dott   & -0.0023  & 0.569    & 0.0016   & 0.524    & -0.0024  & 0.318    & 0.0702   & 0.793    & -0.1403  & 0.628 \\
			 & -0.0012  & 0.709    & 0.0081   & 0.138    & 0.0028   & 0.464    & 0.0000   & 0.993    & -0.2946  & 0.334    & 0.4371   & 0.176 \\
			 \hdashline
	\multirow{5}[0]{*}{Mean PD \footnote{Mean probing depth.}} & 0.0088   & 0.111    & 0.0014   & 0.586    & 0.0052   & 0.032\dottt   & 0.0097   & 0.016\dottt   & -0.2420  & 0.008\dotttt  & -0.3114  & 0.002\dotttt \\
			 & -0.0033  & 0.001\dottttt & 0.0062   & 0.000\dottttt & -0.0006  & 0.796    & -0.0048  & 0.000\dottttt & 0.0878   & 0.409    & -0.1943  & 0.116 \\
			 & 0.0056   & 0.023\dottt   & -0.0031  & 0.140    & 0.0001   & 0.938    & 0.0017   & 0.264    & -0.2412  & 0.037\dottt   & -0.0685  & 0.596 \\
			 & 0.0018   & 0.420    & -0.0043  & 0.046\dottt   & -0.0004  & 0.789    & -0.0010  & 0.425    & 0.2354   & 0.103    & -0.3275  & 0.036\dottt \\
			 & -0.0013  & 0.453    & 0.0034   & 0.242    & 0.0024   & 0.241    & 0.0018   & 0.390    & -0.0132  & 0.936    & 0.1198   & 0.489 \\
			 \hdashline
	\multirow{5}[0]{*}{Mean BP \footnote{Mean bleeding on probing.}} & 0.0069   & 0.108    & 0.0002   & 0.937    & 0.0014   & 0.444    & 0.0030   & 0.341    & -0.3098  & 0.000\dottttt & -0.3442  & 0.000\dottttt \\
			 & -0.0032  & 0.000\dottttt & 0.0053   & 0.000\dottttt & 0.0018   & 0.285    & -0.0030  & 0.002\dotttt  & 0.0462   & 0.575    & -0.0327  & 0.733 \\
			 & 0.0049   & 0.010\dotttt  & 0.0011   & 0.491    & 0.0000   & 0.964    & 0.0007   & 0.591    & -0.0168  & 0.851    & 0.0621   & 0.535 \\
			 & 0.0023   & 0.191    & -0.0025  & 0.138    & 0.0025   & 0.017*   & 0.0000   & 0.975    & 0.1172   & 0.292    & -0.1597  & 0.186 \\
			 & -0.0016  & 0.224    & 0.0039   & 0.080\dott   & 0.0024   & 0.130    & 0.0014   & 0.405    & 0.0612   & 0.629    & -0.0394  & 0.769 \\
	\bottomrule
	\end{tabularx}\vspace{-5pt}
\end{minipage}
\end{sidewaystable}

\begin{table}
	\caption{The loading values and ranks of potentially clinically meaningful taxa in ZIPFA and log-SVD}\label{tab:loading}
\vspace{1ex}
\hspace{-5ex}\begin{tabular}{lcclcc}
	\toprule
	\multicolumn{3}{c}{ZIPFA} & \multicolumn{3}{c}{log-SVD} \\
	\multicolumn{1}{c}{taxa} & rank & loading & \multicolumn{1}{c}{taxa} & rank & loading\\
	\midrule
	Actinomyces\_sp\_oral\_taxon\_169 & 1   & -0.245 & Eubacterium[11][G-3]\_brachy & 10  & 0.114 \\
	Streptococcus\_mutans & 3   & 0.187 & Filifactor\_alocis & 13  & 0.111 \\
	Eubacterium[11][G-5]\_saphenum & 5   & 0.172 & Streptococcus\_sanguinis & 18  & 0.107 \\
	Treponema\_Genus\_probe\_1 & 7   & 0.171 & Eubacterium[11][G-5]\_saphenum & 29  & 0.088 \\
	Prevotella\_melaninogenica & 8   & 0.152 & Treponema\_Genus\_probe\_4 & 30  & 0.086 \\
	Treponema\_Genus\_probe\_4 & 9   & 0.142 & Rothia\_aeria & 35  & -0.084 \\
	Porphyromonas\_gingivalis & 11  & 0.123 & Eubacterium[11][G-6]\_nodatum & 37  & 0.083 \\
	TM7[G-1]\_sp\_oral\_taxon\_346 & 15  & 0.113 & TM7[G-1]\_sp\_oral\_taxon\_346 & 38  & 0.083 \\
	Porphyromonas\_Genus\_probe\_2 & 21  & 0.101 & TM7\_Genus\_probe & 44  & 0.078 \\
	Filifactor\_Genus\_probe & 25  & 0.097 & Prevotella\_nigrescens & 47  & 0.074 \\
	Eubacterium[11][G-6]\_nodatum & 26  & 0.092 & Prevotella\_baroniae & 48  & 0.073 \\
	Filifactor\_alocis & 27  & 0.090 & Filifactor\_Genus\_probe & 58  & 0.068 \\
	Atopobium\_parvulum & 28  & 0.086 & Treponema\_Genus\_probe\_1 & 60  & 0.066 \\
	Prevotella\_Genus\_probe\_2 & 34  & 0.079 & Prevotella\_Genus\_probe\_2 & 62  & 0.065 \\
	Prevotella\_baroniae & 39  & 0.074 & Porphyromonas\_gingivalis & 82  & 0.053 \\
	TM7\_Genus\_probe & 41  & 0.071 & Porphyromonas\_Genus\_probe\_2 & 84  & 0.053 \\
	Actinomyces\_sp\_oral\_taxon\_175 & 58  & -0.062 & Prevotella\_intermedia & 89  & 0.051 \\
	Streptococcus\_sanguinis & 60  & 0.060 & Atopobium\_rimae & 97  & 0.047 \\
	Atopobium\_rimae & 67  & 0.057 & Prevotella\_melaninogenica & 104 & 0.044 \\
	Prevotella\_intermedia & 108 & 0.037 & Actinomyces\_naeslundii & 122 & -0.034 \\
	Rothia\_aeria & 132 & -0.026 & Actinomyces\_sp\_oral\_taxon\_170 & 123 & -0.034 \\
	Eubacterium[11][G-3]\_brachy & 134 & 0.025 & Actinomyces\_sp\_oral\_taxon\_169 & 140 & -0.028 \\
	Actinomyces\_naeslundii & 142 & -0.023 & Streptococcus\_mutans & 147 & 0.025 \\
	Actinomyces\_sp\_oral\_taxon\_170 & 161 & -0.019 & Actinomyces\_sp\_oral\_taxon\_175 & 195 & -0.011 \\
	Prevotella\_nigrescens & 206 & 0.006 & Atopobium\_parvulum & 196 & 0.010 \\
	\bottomrule
	\end{tabular}%
\end{table}

\clearpage
\newpage
\bibliographystyle{biom}  \bibliography{ZIPFA}

\begin{thebibliography}{}

\bibitem[\protect\citeauthoryear{Abusleme, Dupuy, Dutzan, Silva, Burleson,
  Strausbaugh, Gamonal, and Diaz}{Abusleme
  et~al.}{2013}]{abusleme2013subgingival}
Abusleme, L., Dupuy, A.~K., Dutzan, N., Silva, N., Burleson, J.~A.,
  Strausbaugh, L.~D., Gamonal, J., and Diaz, P.~I. (2013).
\newblock The subgingival microbiome in health and periodontitis and its
  relationship with community biomass and inflammation.
\newblock {\em The ISME Journal} {\bf 7,} 1016--1025.

\bibitem[\protect\citeauthoryear{Borsanelli, Gaetti-Jardim~Jr, Schweitzer,
  Viora, Busin, Riggio, and Dutra}{Borsanelli
  et~al.}{2017}]{borsanelli2017black}
Borsanelli, A.~C., Gaetti-Jardim~Jr, E., Schweitzer, C.~M., Viora, L., Busin,
  V., Riggio, M.~P., and Dutra, I.~S. (2017).
\newblock Black-pigmented anaerobic bacteria associated with ovine
  periodontitis.
\newblock {\em Veterinary Microbiology} {\bf 203,} 271--274.

\bibitem[\protect\citeauthoryear{Bray and Curtis}{Bray and
  Curtis}{1957}]{doi:10.2307/1942268}
Bray, R.~J. and Curtis, T.~J. (1957).
\newblock An ordination of the upland forest communities of southern wisconsin.
\newblock {\em Ecological Monographs} {\bf 27,} 325--349.

\bibitem[\protect\citeauthoryear{Cao, Zhang, and Li}{Cao
  et~al.}{2017}]{cao2017microbial}
Cao, Y., Zhang, A., and Li, H. (2017).
\newblock Microbial composition estimation from sparse count data.
\newblock {\em arXiv preprint arXiv:1706.02380} .

\bibitem[\protect\citeauthoryear{Dani, Prabhu, Chaitra, Desai, Patil, and
  Rajeev}{Dani et~al.}{2016}]{dani2016assessment}
Dani, S., Prabhu, A., Chaitra, K., Desai, N., Patil, S.~R., and Rajeev, R.
  (2016).
\newblock Assessment of streptococcus mutans in healthy versus gingivitis and
  chronic periodontitis: A clinico-microbiological study.
\newblock {\em Contemporary Clinical Dentistry} {\bf 7,} 529--534.

\bibitem[\protect\citeauthoryear{Demmer, Jacobs~Jr, Singh, Zuk, Rosenbaum,
  Papapanou, and Desvarieux}{Demmer et~al.}{2015}]{demmer2015periodontal}
Demmer, R., Jacobs~Jr, D., Singh, R., Zuk, A., Rosenbaum, M., Papapanou, P.,
  and Desvarieux, M. (2015).
\newblock Periodontal bacteria and prediabetes prevalence in origins: the oral
  infections, glucose intolerance, and insulin resistance study.
\newblock {\em Journal of Dental Research} {\bf 94,} 201S--211S.

\bibitem[\protect\citeauthoryear{Demmer, Breskin, Rosenbaum, Zuk, LeDuc,
  Leibel, Paster, Desvarieux, Jacobs~Jr, and Papapanou}{Demmer
  et~al.}{2017}]{demmer2017subgingival}
Demmer, R.~T., Breskin, A., Rosenbaum, M., Zuk, A., LeDuc, C., Leibel, R.,
  Paster, B., Desvarieux, M., Jacobs~Jr, D.~R., and Papapanou, P.~N. (2017).
\newblock The subgingival microbiome, systemic inflammation and insulin
  resistance: the oral infections, glucose intolerance and insulin resistance
  study.
\newblock {\em Journal of Clinical Periodontology} {\bf 44,} 255--265.

\bibitem[\protect\citeauthoryear{Dewhirst, Chen, Izard, Paster, Tanner, Yu,
  Lakshmanan, and Wade}{Dewhirst et~al.}{2010}]{dewhirst2010human}
Dewhirst, F.~E., Chen, T., Izard, J., Paster, B.~J., Tanner, A.~C., Yu, W.-H.,
  Lakshmanan, A., and Wade, W.~G. (2010).
\newblock The human oral microbiome.
\newblock {\em Journal of Bacteriology} {\bf 192,} 5002--5017.

\bibitem[\protect\citeauthoryear{Eisen, Spellman, Brown, and Botstein}{Eisen
  et~al.}{1998}]{eisen1998cluster}
Eisen, M.~B., Spellman, P.~T., Brown, P.~O., and Botstein, D. (1998).
\newblock Cluster analysis and display of genome-wide expression patterns.
\newblock {\em Proceedings of the National Academy of Sciences} {\bf 95,}
  14863--14868.

\bibitem[\protect\citeauthoryear{Griffen, Beall, Campbell, Firestone, Kumar,
  Yang, Podar, and Leys}{Griffen et~al.}{2012}]{griffen2012distinct}
Griffen, A.~L., Beall, C.~J., Campbell, J.~H., Firestone, N.~D., Kumar, P.~S.,
  Yang, Z.~K., Podar, M., and Leys, E.~J. (2012).
\newblock Distinct and complex bacterial profiles in human periodontitis and
  health revealed by 16s pyrosequencing.
\newblock {\em The ISME journal} {\bf 6,} 1176--1185.

\bibitem[\protect\citeauthoryear{Hamady and Knight}{Hamady and
  Knight}{2009}]{hamady2009microbial}
Hamady, M. and Knight, R. (2009).
\newblock Microbial community profiling for human microbiome projects: Tools,
  techniques, and challenges.
\newblock {\em Genome Research} {\bf 19,} 1141--1152.

\bibitem[\protect\citeauthoryear{Kumar, Griffen, Barton, Paster, Moeschberger,
  and Leys}{Kumar et~al.}{2003}]{kumar2003new}
Kumar, P., Griffen, A., Barton, J., Paster, B., Moeschberger, M., and Leys, E.
  (2003).
\newblock New bacterial species associated with chronic periodontitis.
\newblock {\em Journal of Dental Research} {\bf 82,} 338--344.

\bibitem[\protect\citeauthoryear{Lambert}{Lambert}{1992}]{lambert1992zero}
Lambert, D. (1992).
\newblock Zero-inflated poisson regression, with an application to defects in
  manufacturing.
\newblock {\em Technometrics} {\bf 34,} 1--14.

\bibitem[\protect\citeauthoryear{Lee, Chugh, Shen, Eberle, and Dittmer}{Lee
  et~al.}{2013}]{lee2013poisson}
Lee, S., Chugh, P.~E., Shen, H., Eberle, R., and Dittmer, D.~P. (2013).
\newblock Poisson factor models with applications to non-normalized microrna
  profiling.
\newblock {\em Bioinformatics} {\bf 29,} 1105--1111.

\bibitem[\protect\citeauthoryear{Li, Huang, and Shen}{Li
  et~al.}{2018}]{li2018exponential}
Li, G., Huang, J.~Z., and Shen, H. (2018).
\newblock Exponential family functional data analysis via a low-rank model.
\newblock {\em Biometrics} {\bf 74,} 1301--1310.

\bibitem[\protect\citeauthoryear{Li}{Li}{2015}]{li2015microbiome}
Li, H. (2015).
\newblock Microbiome, metagenomics, and high-dimensional compositional data
  analysis.
\newblock {\em Annual Review of Statistics and Its Application} {\bf 2,}
  73--94.

\bibitem[\protect\citeauthoryear{Lombardo~Bedran, Marcantonio, Spin~Neto,
  Alves~Mayer, Grenier, Spolidorio, and Spolidorio}{Lombardo~Bedran
  et~al.}{2012}]{lombardo2012porphyromonas}
Lombardo~Bedran, T.~B., Marcantonio, R. A.~C., Spin~Neto, R., Alves~Mayer,
  M.~P., Grenier, D., Spolidorio, L.~C., and Spolidorio, D.~P. (2012).
\newblock Porphyromonas endodontalis in chronic periodontitis: a clinical and
  microbiological cross-sectional study.
\newblock {\em Journal of Oral Microbiology} {\bf 4,} 10123.

\bibitem[\protect\citeauthoryear{Lu, Tomfohr, and Kepler}{Lu
  et~al.}{2005}]{lu2005identifying}
Lu, J., Tomfohr, J.~K., and Kepler, T.~B. (2005).
\newblock Identifying differential expression in multiple sage libraries: an
  overdispersed log-linear model approach.
\newblock {\em BMC Bioinformatics} {\bf 6,} 165.

\bibitem[\protect\citeauthoryear{McMurdie and Holmes}{McMurdie and
  Holmes}{2014}]{mcmurdie2014waste}
McMurdie, P.~J. and Holmes, S. (2014).
\newblock Waste not, want not: why rarefying microbiome data is inadmissible.
\newblock {\em PLoS Computational Biology} {\bf 10,} e1003531.

\bibitem[\protect\citeauthoryear{Mor{\'e}}{Mor{\'e}}{1978}]{more1978levenberg}
Mor{\'e}, J.~J. (1978).
\newblock The levenberg-marquardt algorithm: implementation and theory.
\newblock {\em Lecture Notes in Mathematics, Berlin Springer Verlag} {\bf 630,}
  105--116.

\bibitem[\protect\citeauthoryear{Mor{\'e} and Sorensen}{Mor{\'e} and
  Sorensen}{1983}]{more1983computing}
Mor{\'e}, J.~J. and Sorensen, D.~C. (1983).
\newblock Computing a trust region step.
\newblock {\em SIAM Journal on Scientific and Statistical Computing} {\bf 4,}
  553--572.

\bibitem[\protect\citeauthoryear{Socransky, Haffajee, Cugini, Smith, and
  Kent~Jr}{Socransky et~al.}{1998}]{socransky1998microbial}
Socransky, S., Haffajee, A., Cugini, M., Smith, C., and Kent~Jr, R. (1998).
\newblock Microbial complexes in subgingival plaque.
\newblock {\em Journal of Clinical Periodontology} {\bf 25,} 134--144.

\bibitem[\protect\citeauthoryear{Sohn and Li}{Sohn and Li}{2018}]{b2018glm}
Sohn, M.~B. and Li, H. (2018).
\newblock A glm-based latent variable ordination method for microbiome samples.
\newblock {\em Biometrics} {\bf 74,} 448--457.

\bibitem[\protect\citeauthoryear{Srivastava and Chen}{Srivastava and
  Chen}{2010}]{srivastava2010two}
Srivastava, S. and Chen, L. (2010).
\newblock A two-parameter generalized poisson model to improve the analysis of
  rna-seq data.
\newblock {\em Nucleic Acids Research} {\bf 38,} e170--e170.

\bibitem[\protect\citeauthoryear{Wade}{Wade}{1996}]{wade1996role}
Wade, W. (1996).
\newblock The role of eubacterium species in periodontal disease and other oral
  infections.
\newblock {\em Microbial Ecology in Health and Disease} {\bf 9,} 367--370.

\bibitem[\protect\citeauthoryear{Xu, Paterson, Turpin, and Xu}{Xu
  et~al.}{2015}]{xu2015assessment}
Xu, L., Paterson, A.~D., Turpin, W., and Xu, W. (2015).
\newblock Assessment and selection of competing models for zero-inflated
  microbiome data.
\newblock {\em PLOS ONE} {\bf 10,} e0129606.

\bibitem[\protect\citeauthoryear{Yuan}{Yuan}{1999}]{yuan1999nonlinear}
Yuan, Y. (1999).
\newblock Nonlinear optimization: trust region algorithms.
\newblock In {\em Proceedings of the Third Chinese SIAM Conference}, pages
  84--102.

\end{thebibliography}
\end{document}